\documentclass[twocolumn, prx, aps, superscriptaddress, longbibliography]{revtex4-2}
\usepackage{graphicx}
\usepackage{amsmath}
\usepackage{bm}
\usepackage{hyperref}
\usepackage{bbold}
\newcommand{\sruo}{{Sr$_2$RuO$_4$}}
\usepackage{pdfpages} 
\usepackage{pgffor} 

\makeatletter
\AtBeginDocument{\let\LS@rot\@undefined}
\makeatother

\def\supplementfilename{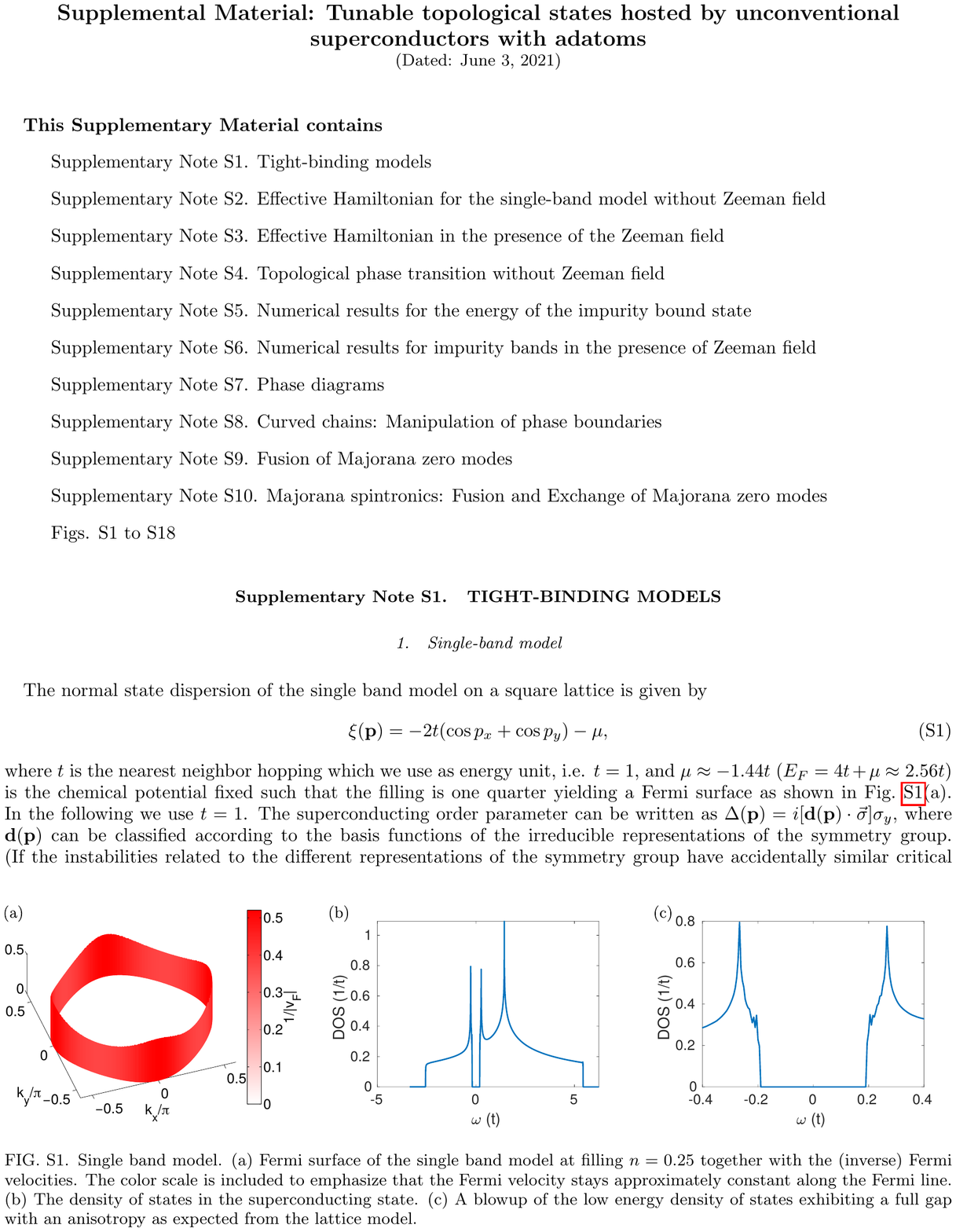}

\pdfximage{\supplementfilename}
\def\numbersupplementpages{\the\pdflastximagepages}

\newif\ifarXiv
\arXivtrue 

\begin{document}
\title{Tunable topological states hosted by unconventional superconductors with adatoms}

\author{Andreas Kreisel}
\affiliation{Institut f\"ur Theoretische Physik, Universit\"at Leipzig, Br\"uderstrasse 16, 04103 Leipzig, Germany}

\author{Timo Hyart}
\affiliation{International Research Centre MagTop, Institute of Physics, Polish Academy of Sciences, Aleja Lotnikow 32/46, PL-02668 Warsaw, Poland}
\affiliation{Department of Applied Physics, Aalto University, 00076 Aalto, Espoo, Finland}

\author{Bernd Rosenow}
\affiliation{Institut f\"ur Theoretische Physik, Universit\"at Leipzig, Br\"uderstrasse 16, 04103 Leipzig, Germany}

\date{July 15, 2021}
 \begin{abstract} 
{
Chains of magnetic atoms, placed on the surface of s-wave superconductors, have been established as a laboratory for the study of Majorana bound states.
In such systems, the breaking of time reversal due to  magnetic 
moments gives rise 
to the formation of in-gap states, which hybridize to form one-dimensional topological superconductors. However, in unconventional superconductors  even non-magnetic impurities  induce in-gap states
since scattering of Cooper pairs changes their momentum but not their phase.  
Here, we propose a  path
for creating topological superconductivity, which is based on an unconventional superconductor with a chain of non-magnetic adatoms on its surface.
The topological phase can be reached by tuning the magnitude and direction of a Zeeman field,
such that  Majorana zero modes at its boundary can be generated, moved and fused.
To demonstrate the feasibility of this platform, we develop a general mapping of films with adatom chains to one-dimensional lattice Hamiltonians. This allows us to study unconventional superconductors such as Sr$_2$RuO$_4$ exhibiting multiple bands and an anisotropic order parameter. 
}
\end{abstract}

\maketitle

\begin{figure}[bt]
%
\begin{center}
 \includegraphics[width=\linewidth]{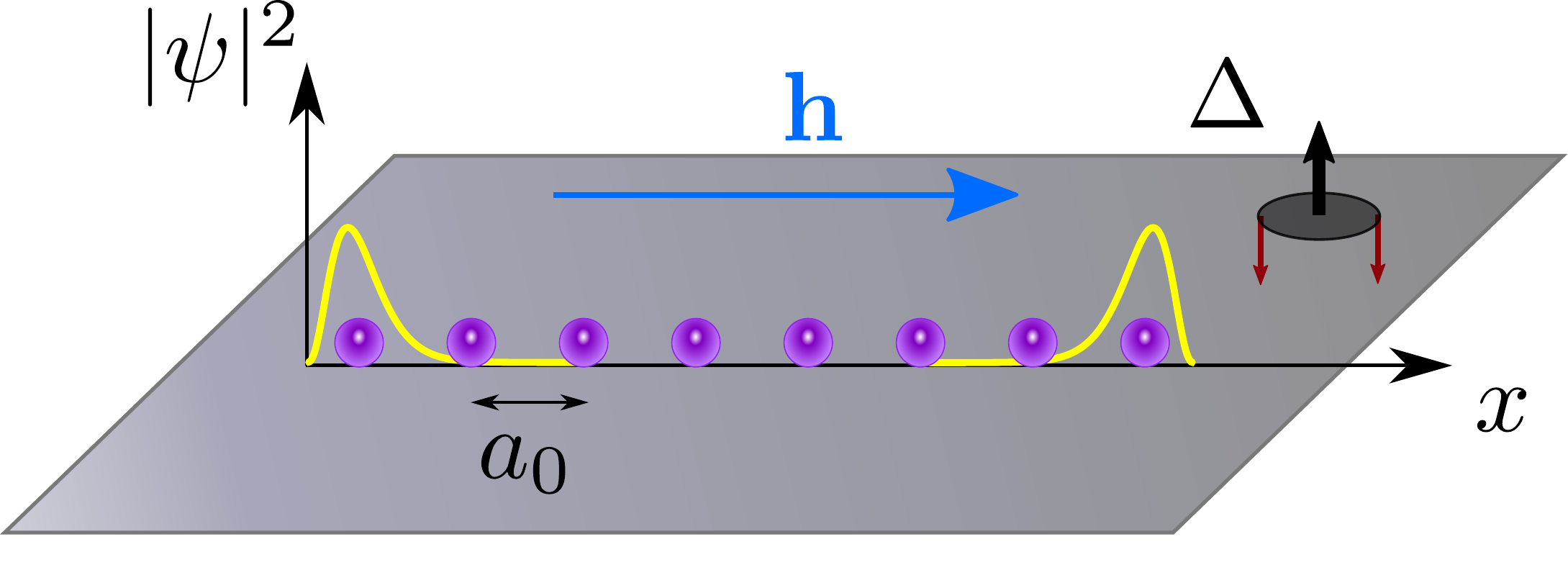}
\caption{{Setup --} Non-magnetic adatoms (purple balls) are placed at a distance $a_0$ to form a chain on the surface of 
a helical triplet superconductor with order parameter $\Delta$.  The Cooper pairs exhibit equal 
spin (red arrows), and their orbital angular momentum (black arrow) points opposite to the spin direction.
An external Zeeman field $\mathbf h$  can be used to tune the system
into the topological phase supporting Majorana zero modes at the endpoints of the chain as sketched by the yellow plot of the magnitude $|\psi|^2$ of the wavefunction.}
\label{model_sketch}
\end{center}
\end{figure}
\section{Introduction}
Combining topology and superconductivity has been heralded as a new paradigm for the realization of exotic new particles -- Majorana zero modes (MZMs)
--whose non-Abelian braiding statistics
would enable fault-tolerant quantum computations\cite{Nay08, Sar15}. Moreover, the existence of MZMs
is topologically
protected, making them inert to  disorder effects. 
To-date, two main approaches, based on the Kitaev chain\cite{Kitaev_2001}, have been pursued in the quest for topological superconductors.
In the first approach $s$-wave
superconductivity\, is proximity-induced in nanowires with strong spin-orbit coupling\cite{Lut10,Oreg10,Mou12, Lutchyn18}, 
while in the second approach the hybridization of impurity (Shiba) bound states gives rise to a topologically nontrivial superconducting phase\cite{Choy11,Martin2012, Pientka13, Brydon15, Kimme16, Ojanen17}. Experimentally, the latter has been realized by placing 
a chain of magnetic atoms
on the surface of an $s$-wave superconductor\cite{Nad13, Nad14,Ruby15,Paw16,Kim18}.
In these platforms, evidences for MZMs at the end points of the system were found in transport measurements on  nanowires\cite{Mou12, Lutchyn18} and in scanning tunneling spectroscopy on Shiba chains\cite{Nad14,Ruby15,Paw16,Kim18,Zhang19}.
While a number of possibilities of chains of addatoms of potential scatterers, magnetic scatterers and nanowires on top of superconductors have been investigated theoretically\cite{Nakosai2013,Kaladzhyan2016,Neupert2016,Kaladzhyan2018,sedlmayr2021new} and proposals for moving, fusing and braiding the MZMs have been brought forward\cite{Alicea2011, Hyart13, Li_Majorana_ring_2016, Aasen_milestones_2016}, it remains an open challenge experimentally to implement those.

In this work, we propose a  path to realize  one-dimensional topological superconductivity 
by placing non-magnetic atoms on the surface of an unconventional triplet superconductor, see Fig.~\ref{model_sketch}. 
The key advantage of our proposal is that it is possible to move and fuse  MZMs by controlling a  magnetic Zeeman field.
Since candidate systems for the realization of triplet superconductivity usually exhibit multiple bands, we go beyond a single-band description and use a model for  Sr$_2$RuO$_4$  to demonstrate that our method can easily be applied to multiband superconductivity. At the same time, we note that Sr$_2$RuO$_4$ has been subject to intense theoretical and experimental investigations regarding the nature of the superconducting pairing \cite{Mackenzie2017, kivelson2020proposal, Pustogow19, leggett2020symmetry, chronister2020evidence}.
Implementing our proposal in  candidate systems for  triplet superconductivity 
such as UPt$_3$\cite{Schemm2014,Taillefer_rev_UPt3}, UTe$_2$\cite{Ran2019,Jiao20} and LaNiGa$_2$\cite{Weng2016} could establish a new MZM platform and improve the understanding of the  pairing symmetries in these systems.

\begin{figure*}[tb]
\begin{center}
 \includegraphics[width=\linewidth]{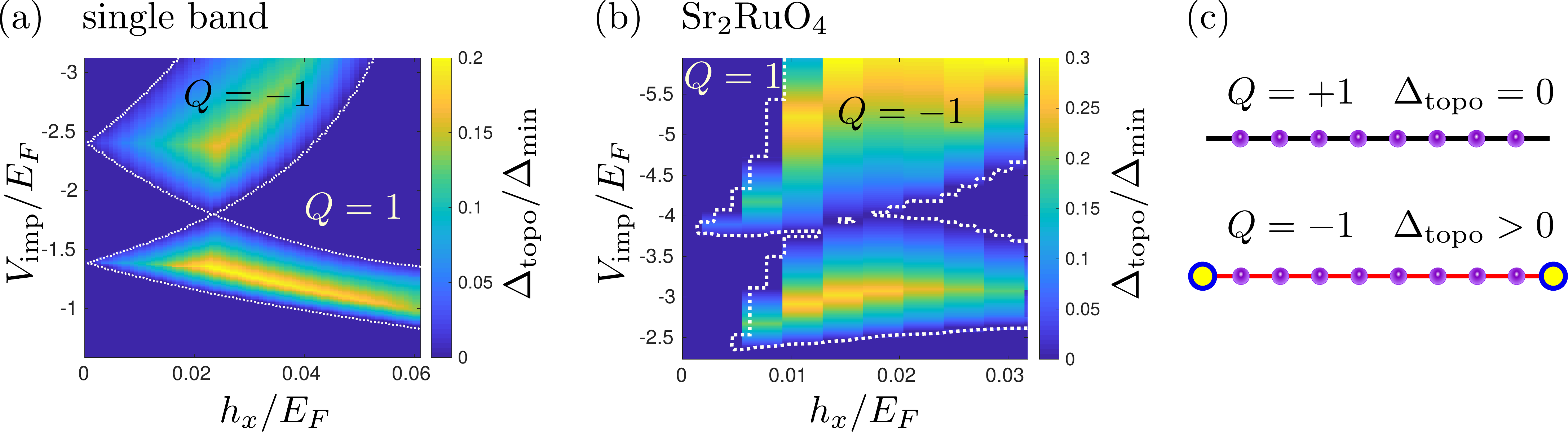}
 \caption{{Topological phase diagram --} (a) Helical single-band superconductor: the topological phase  ($Q=-1$ with finite topological 
 gap $\Delta_{\text{topo}}$) can be reached 
 for a wide range of impurity strengths $V_{\text{imp}}$ by tuning  an external Zeeman field $h_x$.
 The phase boundary is marked with a white dashed line. An energy gap $\Delta_{\text{topo}}\approx \Delta_{\text{min}}/5$ can be reached in the topological phase, leading to well-localized MZMs at the length scale  $\zeta/a_0\approx \Delta_{\text{min}}/\Delta_{\text{topo}}$.
Parameters are: $\Delta_0/E_F\approx 0.08$, $a_0\approx 2.5 \xi$.
(b) Multiband model for Sr$_2$RuO$_4$ with a  realistic superconducting order parameter (see Supplemental material\cite{Note1}) shows similar tunability as the single-band model, with a maximum gap $\Delta_{\text{topo}}\approx \Delta_{\text{min}}/3$.
 (c) For the topological invariant $Q=+1$ the system is in the trivial phase (and we denote $\Delta_{\rm topo}=0$), while for $Q=-1$ MZMs at the ends of the impurity chain exist and are protected against perturbations of the order of the topological gap
 $\Delta_{\text{topo}}$. }
\label{topo_phase_diagram}
\end{center}
\end{figure*}

\section{Model}
\subsection{Bulk superconductor}
Our starting point is a single-band Hamiltonian
\begin{equation}
 H=H_{\mathrm{BdG}}+ H_{\mathrm{Z}}
 \label{eq_full_Hamiltonian}
\end{equation}
where $H_{\mathrm{BdG}}$ describes  a bulk triplet superconductor on a two dimensional lattice, and $H_{\mathrm{Z}}$ is the Zeeman term 
in an external field. In momentum space it has the matrix structure
\begin{equation}
 H(\mathbf {p})=\left(
 \begin{array}{cc}
                                       h(\mathbf {p})  & \Delta(\mathbf {p})\\
                                       \Delta(\mathbf {p})^\dagger& -h(-\mathbf {p})^T
                                      \end{array}
                                      \right)
                                      \label{eq_h_bulk},
\end{equation}
where $h(\mathbf {p}) = \xi(\mathbf {p})\cdot \sigma_0 + \mathbf{h} \cdot {\pmb\sigma}$,
$\xi(\mathbf {p})$  is the energy-momentum dispersion, $\mathbf{h}$  the Zeeman field, and $\sigma_0$ and ${\pmb\sigma}$ are the unit matrix and Pauli matrices acting in  spin space. In the following, we measure all energies in units of the Fermi energy $E_F$.
In the off-diagonal, $\Delta(\mathbf {p})=i[\mathbf{d}(\mathbf{p})\cdot \bm{\sigma}] \sigma_y$ is the pairing term, whose minimum value we denote by $\Delta_{\rm min}$; for details see Appendix \ref{ap_tight_bind}. Here, 
 $\mathbf {d}(\mathbf{p})$ describes the vector order parameter of the triplet superconductor. Here, we concentrate on the case of a helical $p$-wave order parameter 
\begin{align}
 \mathbf {d}_h=-i\Delta_t (\mathbf e_x \sin p_y+\mathbf e_y \sin p_x)\, ,
 \label{eq_helical}
\end{align}
and in the Supplemental information we discuss the generalizations to a chiral $p$-wave order parameter and multiband models using Sr$_2$RuO$_4$ as an example.

\subsection{Chain of nonmagnetic impurities}
The chain of atoms placed along 
the $x$-direction  $\mathbf{r}_n=na_0 \mathbf{e}_x$ (with integer $n$) is 
described by 
\begin{equation}
 H_{\mathrm{imp}}=\hat U \sum_n \delta_{\mathbf{r},\mathbf{r}_n},
 \label{eq_H_imp}
\end{equation}
with the matrix $\hat U=V_{\mathrm{imp}}\tau_z\sigma_0$ mediating non-magnetic impurity scattering of strength $V_{\mathrm{imp}}$. Although the scattering of  Bogoliubov quasiparticles preserves  spin, 
there are still Shiba in-gap states in this system because  the scattering does not change the phase in order to match the $p$-wave momentum dependence of the order parameter.  In the case of a chain,  Shiba states localized in the vicinity of the impurity atoms hybridize and give rise to impurity bands within the bulk gap $\Delta_{\text{min}}$ of the superconductor. These 
bands can be accurately described by an effective Hamiltonian
\begin{equation}
 H_{\mathrm{eff}}(k_x)=\hat U^{-1} \tilde G^{-1} [G_I(k_x)\hat U -1] \ ,
 \label{eq_eff_H}
\end{equation}
which depends on the momentum  $k_x$ in a supercell Brillouin zone with lattice constant $a_0$. We derive the matrices on the r.h.s.
by linearizing the bulk Green function 
with respect to energy, 
 and obtain  $G_I(k_x)$, which describes the propagation of Bogoliubov quasiparticles between the impurities, by  Fourier transforming the bulk Green function at the impurity sites. The matrix $\tilde G^{-1}$  enters as a prefactor and contains the  renormalization of the bandwidth (see Appendix \ref{ap_green}). 

\section{Effective Hamiltonian}
The mapping onto the effective Hamiltonian Eq.~(\ref{eq_eff_H}) can be understood as integrating out the quantum numbers $p_y$ of momenta perpendicular to the chain. Therefore, in the absence of a Zeeman field the effective Hamiltonian has the structure
\begin{equation}
 H_{\mathrm{eff}}^0(k_x)=\xi_{\mathrm{eff}}(k_x)\tau_z\sigma_0 +\Delta_{\mathrm{eff}}(k_x)\tau_x \sigma_0 \ ,
 \label{eq_Heff_0}
\end{equation}
diagonal in spin space, and describing  twofold degenerate impurity bands inside the bulk superconducting gap.  
Here,  $\xi_{\mathrm{eff}}(k_x)$ and $\Delta_{\mathrm{eff}}(k_x)$
are the effective dispersion and pairing for the impurity chain,  containing further-neighbor coupling terms between the impurity states.

The Hamiltonian $H_{\mathrm{eff}}^0(k_x)$ is time-reversal symmetric, so that the system can support a topological phase with an even number of MZMs at each end of the chain\cite{Kimme2015}. We can  realize unpaired MZMs by 
additionally breaking the symmetry down to only particle-hole symmetry by use of a Zeeman field such that the system, when tuned into the topological phase, 
exhibits unpaired MZMs at the end of the impurity chain
A Zeeman field in $z$-direction can be described by
$H^{\mathrm Z}_{\text{eff},z}=h_{\text{eff},z}\sigma_z\tau_z$  with a renormalized magnitude  $h_{\text{eff},z}$, which  can be calculated by including  chemical potential shifts $\xi(\mathbf p)\pm h_z$ in the bulk dispersion of the spin up and down electrons, respectively. 
\begin{figure*}[tb]
\begin{center}
 \includegraphics[width=\linewidth]{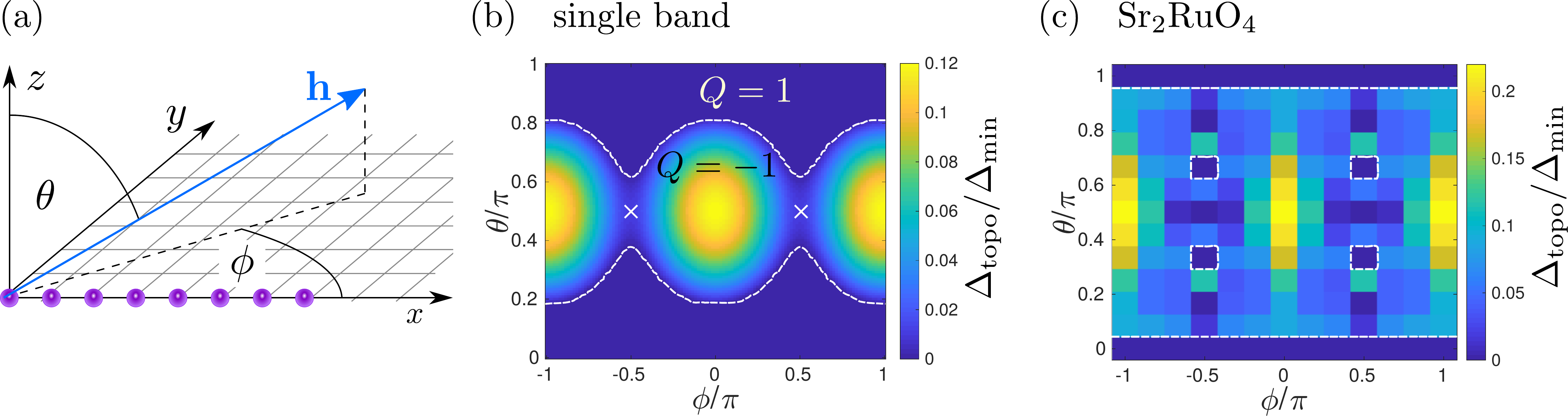}
 \caption{{Field direction dependence of the topological gap --}
 (a) The direction of the Zeeman field relative to the direction of the impurity chain is parametrized by the azimuthal angle $\phi$ and the polar angle  $\theta$. Changing the direction of the field  allows  to enter or leave
 the topological phase. 
 (b) Topological phase diagram for a single-band system with parameters
$V_{\text{imp}}/E_F=-1.2$, $|h|/E_F=0.03$, $a_0\approx 2.5 \xi$; the white dashed line is the boundary between topological and trivial phase. 
 (c) Similar phase diagram for Sr$_2$RuO$_4$
($V_{\text{imp}}/E_F=-5.2$, $|h|/E_F=0.015$, $a_0\approx 0.8 \xi$).
In both cases, the topological gap $\Delta_{\mathrm{topo}}$ is maximal for the field along the impurity chain ($\phi=0$, $\theta=\pi/2$), and the topologically trivial phase can be reached by either tuning to $\phi=\pi/2$,  
or towards $\theta=0, \pi$, i.e. to a transverse directions relative to the chain. The isolated point at which the system remains gapless (white cross) corresponds to a in plane field perpendicular to the chain.}
\label{angle_phase_diagram}
\end{center}
\end{figure*}
We find that  $h_{\text{eff},z}\ll h_z$ because the energy of  Shiba in-gap state depends only weakly on the chemical potential (see Supplemental information). Thus, the effective Hamiltonian is diagonal in spin space with each block $\tilde H_{\mathrm{eff}\pm}=(\xi_\mathrm{eff}(k_x)\pm h_{\text{eff},z})\tau_z +\Delta_{\mathrm{eff}}(k_x)\tau_x$ exhibiting the same topological properties as the Kitaev chain. The effective field $\pm h_{\text{eff},z}$ plays the role of the chemical potential, which  can drive a topological phase transition. 
However, since the effective field $h_{\text{eff},z}$ is parametrically small,  the topological phase can only be reached 
by fine tuning the impurity strength $V_{\rm imp}$ such that the impurity bands  almost touch zero already without a Zeeman field.

For the Zeeman field pointing in $y$-direction, the additional term to the effective Hamiltonian is $H^{\mathrm Z}_{\text{eff},y}=h_{\text{eff},y}\sigma_y\tau_0$.
In this case, the two BdG bands are shifted trivially in energy with respect to each other, leaving their topological character unchanged, i.e.~a field in $y$-direction cannot tune into the topological phase.

Finally, for a field in $x$-direction, the Zeeman term reads $H^{\mathrm Z}_{\text{eff},x}=h_{\text{eff},x}\tau_z\sigma_x$  with weak renormalization of the effective Zeeman field from the  bulk value $h_{\text{eff},x} \sim h_x$.
The effective Hamiltonian 
can now be rotated around the $y$ axis in spin space $\sigma_x\rightarrow \sigma_z$, such that in the new basis the effective field $h_{\text{eff},x}$ plays again the role of a chemical potential.  The difference to the case of a Zeeman field in $z$-direction is that the weakly  renormalized $h_{\text{eff},x}$  can drive a topological phase transition much more efficiently than  the strongly reduced $h_{\text{eff},z}$ discussed above.

\section{Topological phase diagram}
To quantitatively demonstrate the tunability of topological superconductivity, we compute the topological phase diagram (see Figs.~\ref{topo_phase_diagram} and \ref{angle_phase_diagram}) as characterized by the topological invariant 
\begin{equation}
 Q=\prod_{k_x \in \mathrm{TRIM}} Q(k_x),\, \ Q(k_x)=\mathrm{sign} (\mathrm{Pf}[H(k_x)\tau_x]).
 \label{eq_Q_topo}
\end{equation}
It is given as the product of Pfaffians $Q(k_x)$ at the time reversal invariant momenta of the corresponding one-dimensional Brillouin zone, details on its derivation can be found in Appendix \ref{ap_topo_inv}. 
We  supplement the fully numerical supercell calculation by computing the topological invariant also using $Q_{\mathrm{eff}}(k_x)=\mathrm{sign} (\mathrm{Pf}[H_{\mathrm{eff}}(k_x)\tau_x])$. The excellent agreement between  $Q_{\rm eff}$ and $Q$ (see Supplemental Material) indicates that  $H_{\mathrm{eff}}(k_x)$ indeed faithfully describes the low-energy physics of the impurity chain. In the nontrivial case  $Q=-1$ we define the topological gap $\Delta_{\mathrm{topo}}$ as the minimum of the eigenenergies of $H(k_x)$ in the Brillouin zone, see Appendix \ref{ap_topo_gap}.
To detect the non-Abelian properties of the MZMs, it is necessary that 
the coupling between MZMs is  weak. 
Thus,  the distance between neighboring MZMs needs to be much larger than 
 $\zeta \approx \hbar v_{F,\text{eff}}/\Delta_{\mathrm{topo}}$, where $v_{F,\text{eff}}$ is the Fermi velocity for the impurity band.
An estimate of $v_{F,\text{eff}}$  yields $\zeta/a_0\approx \Delta_{\text{min}}/\Delta_{\text{topo}}$.
In order to avoid thermal excitations the temperature needs to be smaller than the topological gap, $k_{\mathrm B}T< \Delta_{\mathrm{topo}}$.

In Fig.~\ref{topo_phase_diagram} we present the topological phase diagram for the single-band model and for a multiband description of Sr$_2$RuO$_4$, revealing
that the topological phase can be reached in both cases by application of a Zeeman field $h_x$ in the direction along the impurity chain for a large range of the impurity potential $V_{\mathrm{imp}}$. Experimentally, suitable adatoms can be identified by examining the appearance of in-gap states in the tunneling spectra.
We summarize that our proposal might be feasible in a helical $p$-wave superconductor where impurity adatoms can be controlled experimentally. 
The multiband helical $p$-wave order parameter considered in Fig.~\ref{topo_phase_diagram} is one of the possible candidate pairing symmetries for Sr$_2$RuO$_4$\cite{leggett2020symmetry}. Controllable placing of  adatoms might be facilitated by step edges, and in the Supplemental material\cite{Note1} we show that the topological phase can be reached by application of a Zeeman field also in this case.
In Fig.~\ref{angle_phase_diagram} we illustrate that the direction of the Zeeman field can be used to tune a system into and out of the topological phase. 
Namely, tuning the azimuthal angle $\phi$ has a strong effect on the topological gap $\Delta_{\mathrm{topo}}$. 
A similar analysis of the tunability of a system with chiral $p$-wave order parameter (see Supplemental Material\cite{Note1}) reveals that rotating the Zeeman field within the $x$-$y$-plane has no effect at all. 
Hence, this difference in behavior   could be used to experimentally  diagnose and discriminate the helical $p$-wave 
from a chiral $p$-wave order parameter, an important question for example in the  Sr$_2$RuO$_4$ system\cite{Mackenzie2017, Pustogow19, kivelson2020proposal, leggett2020symmetry}.

\begin{figure*}[tb]
\begin{center}
 \includegraphics[width=\linewidth]{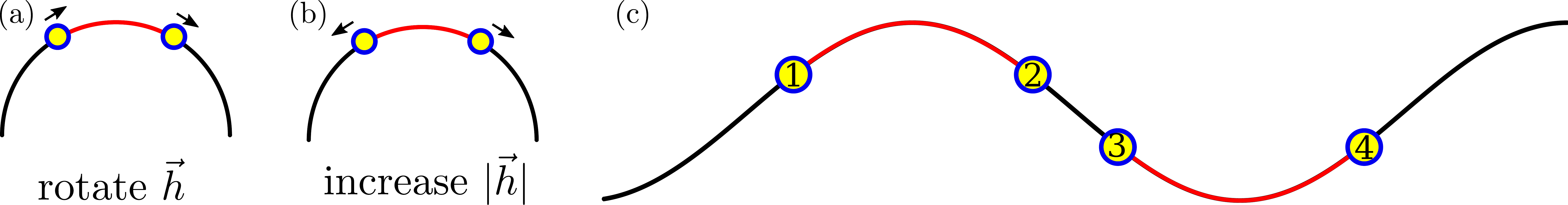}
\caption{{Moving and fusion of Majorana zero modes --}
A curved impurity chain might be tuned partially into the topological state (red arcs) if the  the angle between the Zeeman field and the local direction of the chain puts it into the topological phase. MZMs (yellow dots) occur at the boundaries of trivial and nontrivial regions, which can be moved in two different ways: (a) Rotating the field around the axis perpendicular to the plane moves the Majoranas in the same direction, while (b)  changing the field magnitude or polar angle can move them in opposite directions. (c) In a wiggly impurity chain  MZMs 1 and 2, 3 and 4 can be created pairwise from the vacuum by changing the field magnitude, and by tuning deeper into the topological phase one eventually fuses the MZMs 2 and 3.}
\label{topo2_curved}
\end{center}
\end{figure*}

\section{Discussion}
For the case of magnetic adatoms, the dependence of adatom magnetic order on an external Zeeman field has been suggested as a means to  to create, braid, and fuse MZMs\cite{Li_Majorana_ring_2016}. Here, we exploit the direct dependence of the topological phase diagram on the Zeeman field (see Fig.~3). So far,  we have arbitrarily chosen that the impurity chain is oriented along the $x$-axis, and as a result a Zeeman field in $x$-direction was most suitable to induce a topological phase. More generally however, the relevant parameter is the relative angle of the Zeeman field with the impurity chain, and 
 for a curved  chain
 the relevant angle would be the angle between the field and the local tangential direction as  defined in Fig.~\ref{angle_phase_diagram}(a). Thus, 
 MZMs on curved impurity chains are located at all interfaces between trivial and nontrivial regions, i.e. at  positions where the local tangent and the external field draw a critical angle, see Fig.~\ref{topo2_curved}.
Hence, MZMs can be moved along the impurity chain by either rotating the direction of the Zeeman field  or by 
changing the magnitude of the field,  which modifies the critical angle.  

Increasing the Zeeman field along a
wiggly impurity chain creates two pairs of Majoranas which can formally be described by  operators $\gamma_i$,  $i=1,2,3,4$, satisfying $\gamma_i=\gamma_i^\dag$ and anticommutation relations   $\{\gamma_i,\gamma_j\}=2 \delta_{ij}$ (for the labeling of MZMs see Fig.~\ref{topo2_curved}(c)).
Grouping the MZMs in pairs of two, we can define  the left and right number operators  $n_l=\frac 12 (1+i\gamma_1\gamma_2)$ and $n_r=\frac 12 (1+i\gamma_3\gamma_4)$ with eigenvalues 0 and 1, and define a Hilbert space spanned by basis states $\left|n_l n_r\right>$. Creating the MZMs from the vacuum, the initial state is given by $\Psi = |00\rangle$ in this basis. 
Tuning deeper into the topological phase, the inner MZMs 2 and 3 will fuse  such that the final state 
will be a statistical mixture of $0$ and $1$ for the operator $n_o=\frac 12 (1+i\gamma_1\gamma_4)$ (see Supplemental material\cite{Note1}). In the  fusion process the projective measurement can be performed by detecting the charge acquired by MZMs 2 and 3 after they have  hybridized \cite{Aasen_milestones_2016}. The movement and projective measurements are the key ingredients for manipulation of MZMs, and they can be realized by controlling external magnetic field as discussed above.
However, we also need to preserve the quantum information stored in the MZMs. In the Supplementary material\cite{Note1} we discuss the corresponding requirements and propose to manipulate the local magnetization by spintronic means to obtain signatures of non-Abelian statistics of MZMs.

In summary, we have shown that topological superconductivity can be realized by placing
nonmagnetic adatoms on the surface of an unconventional superconductor, and that the topological invariant can be controlled with the magnitude and direction of a Zeeman field. Our considerations are
based on a lattice Hamiltonian which can describe materials exhibiting a complex structure of the order parameter
and multiple bands. We have identified the field direction which can most efficiently
tune the system into the topological phase and we have  proposed a scheme to move and fuse MZMs. An experimental
realization of this proposal could become a scalable platform for topological quantum information processing based on the non-Abelian statistics of MZMs.

\section*{Acknowledgements}
The research was partially supported by the Foundation for Polish Science through the IRA Programme co-financed by EU within SG OP. We acknowledge support from Leipzig University for Open Access Publishing.
The code to numerically calculate the spectra and topological invariants discussed in this paper  as well as the data that has been used to generate the plots within this paper is available from the corresponding author upon reasonable request.
B.R. acknowledges support by DFG grant RO 2247/11-1.

\appendix
\section{Tight binding model}
\label{ap_tight_bind}
For the single-band model, we use the normal state dispersion $\xi(\mathbf p)=-2t(\cos p_x +\cos p_y)-\mu$ on a square lattice, where $t$ is the nearest-neighbor hopping and $\mu\approx -1.44 t$ is the chemical potential fixed such that the filling is one quarter and the Fermi energy $E_F \approx 2.56 t$.

The superconducting order parameter can be written in real space as
\begin{equation}
 \Delta=\sum_{ij} \sum_{\alpha \beta} \sum_{\sigma \sigma'} \Delta_{ij,\alpha\beta,\sigma\sigma'} c^\dag_{i,\alpha,\sigma}c^\dag_{j,\beta,\sigma'}+h.c..
\end{equation}
For the single-band system with a triplet order parameter, the coefficients read
$\Delta_{ij,\alpha\beta,\sigma\sigma'}= \Delta_0 {\mathbf d}_{i,j}\cdot{\pmb \sigma} i\sigma_y$
with the Pauli operators ${\pmb\sigma}=[\sigma_x,\sigma_y,\sigma_z]$ and a vector ${\mathbf d}_{i,j}$.
In the main text, we consider the physical consequences of the 
helical $p$-wave order parameter ${\mathbf d}=-i \Delta_t(\sin p_y{\mathbf e}_x+\sin p_x{\mathbf e}_y)$
which in real-space leads to nearest-neighbor pairings
$
\Delta_0=\delta_i\left(\begin{array} {cc}
                         -1&0\\
                         0&1
                        \end{array}\right)
$ 
 with $\delta_i=\mp \Delta_t/2$ for the relative lattice vector $(0, \pm 1)$,
and $
\Delta_0=\delta_i\left(\begin{array} {cc}
                         i&0\\
                         0&i
                        \end{array}\right)
$ 
with $\delta_i=\mp\Delta_t/2$ for the relative lattice vector $(\pm 1,0)$.

The multiband model for Sr$_2$RuO$_4$\cite{ScaffidiPRL} is discussed in the Supplemental material\cite{Note1}.

\section{Green function approach and effective Hamiltonian}
\label{ap_green}
Derivations of effective Hamiltonians for continuum models have been worked out in detail for example for helical Shiba chains in Ref.~\onlinecite{Pientka13} and spinless superconductors\cite{Neupert2016}. Here, we generalize this approach
to multiband lattice models and derive the effective Hamiltonian for the impurity bands as cited in Eq.~(\ref{eq_eff_H}) of the main text. Starting point is the eigenvalue equation of the Bogoliubov de Gennes Hamiltonian (including the impurity chain) $(H_{\mathrm{BdG}}+H_{\mathrm Z}+H_{\mathrm{imp}})\Psi=E\Psi$ with the eigenstate $\Psi$ and eigenenergy $E$.
Next, we introduce the Green function operator of the bulk system
\begin{equation}
G(E)=(E-H_{\mathrm{BdG}}-H_{\mathrm Z})^{-1}
\end{equation}
to obtain a nonlinear eigenproblem 
\begin{equation}
\Psi=G(E)H_{\mathrm{imp}}\Psi\,.
\label{eq_ev_1}
\end{equation}
Evaluating Eq.~(\ref{eq_ev_1}) only at the impurity sites $\mathbf r_m=a_0 m \mathbf e_x$, and using that the impurity Hamiltonian [Eq. (\ref{eq_H_imp})] is diagonal, we obtain
$$\Psi(\mathbf r_m)=\sum_{\mathbf r_n} G(E,\mathbf r_m-\mathbf r_n) \hat U\Psi(\mathbf r_n).$$
Because of the periodicity with respect to translations by the impurity chain lattice vector $a_0 \mathbf e_x$ it is useful to transform this equation to  momentum space (with respect to the supercell)
$\Psi(k_x)=\sum_{m}\Psi(\mathbf r_m) e^{-i k_x a_0 m}$
 such that the eigenvalue equation can be rewritten  as
\begin{equation}
 \Psi(k_x)=\sum_{n} G(E,\mathbf r_n) e^{-ik_x a_0 n} \hat U\Psi(k_x)\,.
 \label{eq_psi_k}
\end{equation}

The real space Green function can be obtained via its Fourier representation
\begin{eqnarray}
G(E,\mathbf r)&=&\frac{1}{\Omega_{\mathrm{BZ}}}\int_{\mathrm{BZ}} d^2 p\, G(E,\mathbf p) e^{i\mathbf p\cdot \mathbf r},  \nonumber \\ 
G(E,\mathbf p)&=&[E-H_{\rm BdG}(\mathbf{p})-H_Z]^{-1},
\label{eq_G_FT}
\end{eqnarray}
where the integral is over the bulk Brillouin zone with momentum space area $\Omega_{BZ}$. By linearizing the bulk Green function at $E=0$, we obtain 
\begin{align}
G(E, \mathbf{p})
=G(0,\mathbf{p})-E\tilde G(\mathbf{p}),
\label{eq_G_expansion_p}
\end{align}
where $G(0, \mathbf{p})=- [H_{\rm BdG}(\mathbf{p})+H_{\mathrm Z}]^{-1}$ and $\tilde G(\mathbf{p})=[H_{\rm BdG}(\mathbf{p})+H_{\mathrm Z}]^{-2}$ exist for fully gapped systems. Furthermore, we keep the linear correction $\propto E$ only in the onsite term of the Green function to obtain
\begin{equation}
G(E,\mathbf r_n)=G(0, \mathbf{r}_n)- E \delta_{\mathbf{r}_n, \mathbf{0}}  \tilde{G}, \label{onsitetilde}
\end{equation}
where 
\begin{equation}
\tilde{G} = \frac{1}{\Omega_{BZ}} \int_{BZ} d^2p \ [H_{\rm BdG}(\mathbf{p}) + H_Z]^{-2}\,. \label{Gtilde}
\end{equation}

We now insert Eqs.~(\ref{eq_G_FT}), (\ref{eq_G_expansion_p}) and (\ref{onsitetilde}) into Eq.~(\ref{eq_psi_k}) and introduce the Fourier transforms with respect to the supercell as 
\begin{equation}
G_I(k_x)=\sum_{n} G(0,\mathbf{r}_n) e^{-ia_0 n k_x} \label{GI}
\end{equation}
to obtain
\begin{equation}
\Psi(k_x) 
=[G_I(k_x)\hat{U} -E\tilde{G} \hat{U} ]\Psi(k_x). \label{lineareigenvalueequationp}
\end{equation}
Rearranging the terms and multiplication with inverse matrices brings the eigenvalue equation in the form 
\begin{align}
H_{\mathrm{eff}}(k_x) \Psi(k_x)=E  \Psi(k_x),
\end{align}
where the effective Hamiltonian is
\begin{equation}
H_{\mathrm{eff}}(k_x)=\hat{U}^{-1} \tilde G^{-1} \big[ G_I(k_x) \hat U-1 \big].
 \end{equation}
This Hamiltonian becomes exact at $E=0$ where the topological phase transition occurs and therefore can be used to determine the phase diagram exactly. 

This approach is general and can be applied to all lattice Hamiltonians. In this work we have applied it to the single-band $p$-wave superconductors and multiband model for Sr$_2$RuO$_4$ \cite{ScaffidiPRL}, but similar theoretical investigations can be performed also for other candidate materials for multiband triplet superconductors  \cite{Schemm2014,Taillefer_rev_UPt3,Ran2019,Jiao20,Weng2016}.

\section{Topological invariant}
\label{ap_topo_inv}
A finite magnetic field breaks time-reversal symmetry such that the remaining symmetry of the Hamiltonian, Eq.~(\ref{eq_full_Hamiltonian}) is particle-hole symmetry, described  by a particle-hole operator $P$ anticommuting with the Hamiltonian, $\{H,P\}=0$. The superconducting pairing has the property $\Delta^T=-\Delta$ and the normal state block is Hermitean, $h^\dagger=h$. Therefore,  $P=\tau_x K$ is the  desired anticommuting operator, where $\tau_x$ is the Pauli matrix in particle-hole space and $K$ the complex conjugation. The parity operator $\hat P=(-1)^{\hat N}$, with $\hat N$ being the particle number operator, commutes with the Hamiltonian $[H,\hat P]=0$, thus there is a common system of eigenstates. Since the parity operator has the eigenvalues $\pm 1$, the ground state of the system is either of odd or even parity. The parity can be calculated by Eq.~(\ref{eq_Q_topo}), where we formally have factorized out a prefactor of $(-1)^n$ because $\mathop{\mathrm{Pf}}(Hi\tau_x)=\mathop{\mathrm{Pf}}(H\tau_x)$ for matrices of size $4n$ with $n$ an integer number. In the fully numerical approach, we set up the Hamiltonian for the superconductor subject to the Zeeman field and including the impurity potential, and use a supercell method to obtain $H(k_x)$ of an (infinite) impurity chain along the $x$ direction. For the calculation of the invariant using Eq.~(\ref{eq_Q_topo}) the supercell Hamiltonian needs to be constructed for the two time-reversal invariant momenta, $k_x=0,\pi/a_0$, corresponding to periodic or antiperiodic boundary conditions. Finally, the Pfaffian is calculated using an efficient numerical algorithm\cite{Wimmer2012}.

The effective Hamiltonian, Eq. (\ref{eq_eff_H}) inherits the symmetries from the bulk Hamiltonian in Eq.~(\ref{eq_h_bulk}), i.e.~it satisfies the particle-hole symmetry $\tau_x  H_{\mathrm{eff}}^*(-k_x)\tau_x=-H_{\mathrm{eff}}(k_x)$ which can be read off from Eq.~(\ref{eq_eff_H}) by using that also the other matrices in the expression obey the same symmetry, e.g.~$\tau_x \hat U^*\tau_x=-\hat U$,  $\tau_x \tilde G^*\tau_x=\tilde G$, $\tau_x  G_I^*(-k_x)\tau_x=-G_I(k_x)$, which can be derived from the original property $\tau_x  [H_{\mathrm{BdG}}(\mathbf p)+H_{\mathrm Z}]^*\tau_x=-[H_{\mathrm{BdG}}(\mathbf p)+H_{\mathrm Z}]$  of the bulk Hamiltonian. Therefore, the effective Hamiltonian can be used to calculate the topological invariant using Eq.~(\ref{eq_Q_topo}), while the numerical effort is greatly reduced because of the small size of the corresponding matrices.

\section{Topological gap}
\label{ap_topo_gap}
For the calculation of the topological gap, i.e. the minimal positive eigenvalue of the supercell Hamiltonian as a function of $k_x$, we calculate the eigenvalues for a grid of a few $k_x$ points between $0$ and $\pi/a_0$, select the $k_x$ with the smallest positive eigenvalue, and then use an iterative procedure to find the smallest positive eigenvalue by a bisection bracketing algorithm
to obtain $\Delta_{\mathrm{topo}}$. This procedure is is implemented for both the supercell and the effective Hamiltonian approach 
 to investigate the reliability of the approximation in deriving Eq.~(\ref{eq_eff_H}), see Supplemental Material. Finding very good agreement, we show in the main text only results stemming from the effective Hamiltonian since the calculation of eigenvalues is orders of magnitude faster once the expansion coefficients, Eqs.~(\ref{Gtilde}) and (\ref{GI}), for $H_{\mathrm{eff}}(k_x)$ have been calculated.
 
\nocite{ScaffidiPRL}
\nocite{Romer2020}
\nocite{Hyart14Zeeman}
\nocite{Kimme2015}
\nocite{Kimme16}
\nocite{Aasen_milestones_2016}
\nocite{Beenakker20}
\nocite{Zhang_PRR2020}
\nocite{Tserkovnyak05, Parkin190, Matsukura2015, PhysRevApplied2015, Tokura2019, spintroRev}
\nocite{Beenakker20, Bonderson08, Karzig17}
 \bibliography{literature_topo.bib}
\ifarXiv
    

\section*{Supplementary material}

# Supplemental Material: Tunable topological states hosted by unconventional superconductors with adatoms

(Dated: June 3, 2021)

**This Supplementary Material contains**

## Supplementary Note S1. TIGHT-BINDING MODELS

### 1. Single-band model

The normal state dispersion of the single band model on a square lattice is given by

$$\xi(\mathbf{p}) = -2t(\cos p_x + \cos p_y) - \mu, \tag{S1}$$

where $t$ is the nearest neighbor hopping which we use as energy unit, i.e. $t = 1$, and $\mu \approx -1.44t$ ($E_F = 4t + \mu \approx 2.56t$) is the chemical potential fixed such that the filling is one quarter yielding a Fermi surface as shown in Fig. S1(a). In the following we use $t = 1$. The superconducting order parameter can be written as $\Delta(\mathbf{p}) = i[\mathbf{d}(\mathbf{p}) \cdot \vec{\sigma}]\sigma_y$, where $\mathbf{d}(\mathbf{p})$ can be classified according to the basis functions of the irreducible representations of the symmetry group. (If the instabilities related to the different representations of the symmetry group have accidentally similar critical

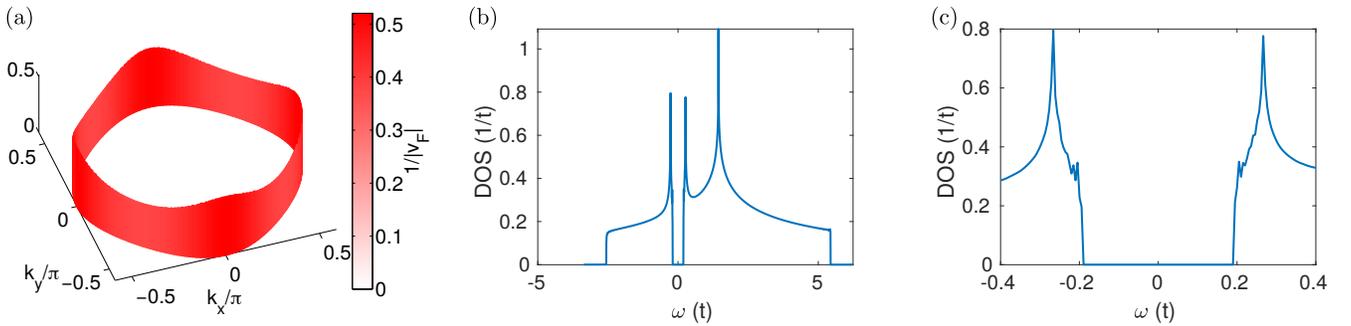

FIG. S1. Single band model. (a) Fermi surface of the single band model at filling $n = 0.25$ together with the (inverse) Fermi velocities. The color scale is included to emphasize that the Fermi velocity stays approximately constant along the Fermi line. (b) The density of states in the superconducting state. (c) A blowup of the low energy density of states exhibiting a full gap with an anisotropy as expected from the lattice model.

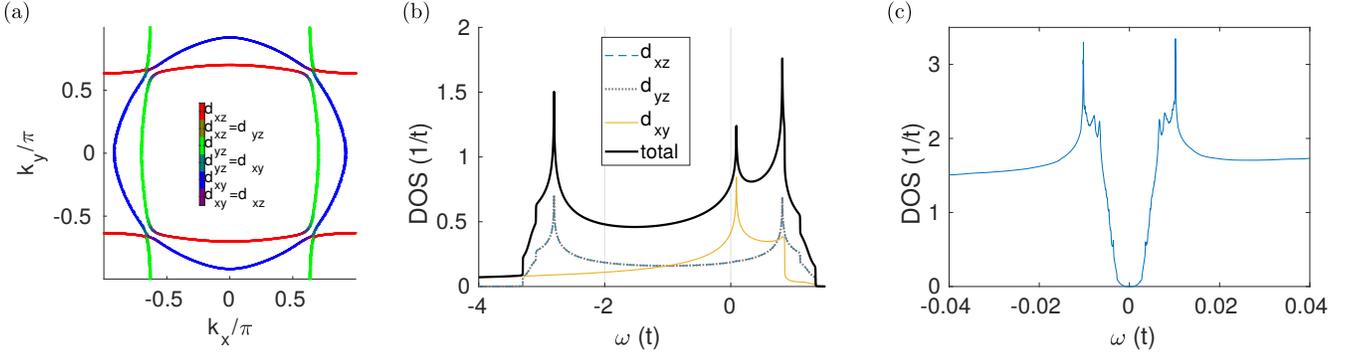

FIG. S2. Three-band model for $Sr_2RuO_4$. (a) Fermi surface including orbital weights. (b) Density of states in the normal state. (c) Density of states in the superconducting state showing a V-shaped feature from the strongly anisotropic order parameter.

temperatures, a second transition to a mixed state can occur upon lowering of the temperature, but this possibility is neglected here.) For the square lattice, the possible triplet order parameters are

$$\mathbf{d}_{1,2}(\mathbf{p}) = \Delta(\mathbf{e}_x \sin p_y \pm \mathbf{e}_y \sin p_x), \quad \mathbf{d}_{3,4}(\mathbf{p}) = \Delta(\mathbf{e}_x \sin p_x \pm \mathbf{e}_y \sin p_y), \tag{S2}$$

$$\mathbf{d}_{5A}(\mathbf{p}) = \Delta_{5A}\mathbf{e}_z \sin p_x, \quad \mathbf{d}_{5B}(\mathbf{p}) = \Delta_{5B}\mathbf{e}_z \sin p_y. \tag{S3}$$

The basis functions of the one-dimensional representations $\mathbf{d}_m(\mathbf{p})$ ($m = 1, 2, 3, 4$) are distinguished by the different signs in the mirror transformations $d_{m,x}(-p_x, p_y) = \pm d_{m,x}(p_x, p_y)$ and $d_{m,x}(p_y, p_x) = \pm d_{m,y}(p_x, p_y)$. In the case of the two-dimensional irreducible representation the basis functions $\mathbf{d}_{5A}(\mathbf{p})$ and $\mathbf{d}_{5B}(\mathbf{p})$ are parallel to $\mathbf{e}_z$.

All basis functions $\mathbf{d}_m(\mathbf{p})$ ($m = 1, 2, 3, 4$) give rise to helical $p$-wave superconductors and from the view point of our analysis they are essentially equivalent. Therefore, we consider

$$\mathbf{d}_h(\mathbf{p}) = -i\Delta_t(\mathbf{e}_x \sin p_y + \mathbf{e}_y \sin p_x) \tag{S4}$$

as a representative example of helical $p$-wave superconductors. If the highest critical temperature corresponds to the two-dimensional irreducible representation the order parameter that is realized (by minimization of the free energy) is a complex linear combination of $\mathbf{d}_{5A}(\mathbf{p})$ and $\mathbf{d}_{5B}(\mathbf{p})$, which describes a chiral $p$-wave superconductor

$$\mathbf{d}_c(\mathbf{p}) = \Delta_t(\sin p_x + i \sin p_y)\mathbf{e}_z. \tag{S5}$$

We choose $\Delta_t = 0.2$ throughout the discussion of the single band model unless stated otherwise. In the single band model with filling $n = 0.25$, the magnitude of the order parameter on the Fermi surface $|\Delta(p)|$ is almost constant, so is the (inverse) Fermi velocity, Fig. S1(a). The density of states is fully suppressed within the energy interval $|\omega| \lesssim 0.2t$ and yields coherence peaks at the gap maxima, see Fig. S1(b,c).

2. *Multiband model for $Sr_2RuO_4$*

The model for $Sr_2RuO_4$ is based on a tight binding parametrization proposed earlier[1] with hoppings on a square lattice giving rise to a Hamiltonian $H(\mathbf{p}) = H_0(\mathbf{p}) + H_{SO}$ with the spin independent part $H_0(\mathbf{p}) = \sum_{ab,s} t_{ab}(\mathbf{p}) c^\dagger_{a,s}(\mathbf{p}) c_{b,s}(\mathbf{p})$ where the sum runs over the Ru-$d$ orbitals $(a, b) = \{d_{xz}, d_{yz}, d_{xy}\}$ and the spin $s = \pm 1$. The Fourier transform of the hopping elements are given by $t_{11}(\mathbf{p}) = -2t \cos p_x - 2t^\perp \cos p_y - \mu$, $t_{22}(\mathbf{p}) = -2t^\perp \cos p_x - 2t \cos p_y - \mu$, $t_{33}(\mathbf{p}) = -2t'[\cos p_x + \cos p_y] - 4t'' \cos p_x \cos p_y - \mu'$, $t_{12}(\mathbf{p}) = t_{21}(\mathbf{p}) = -4t''' \sin p_x \sin p_y$ with $(t, t^\perp, t', t'', t''') = (1.0, 0.1, 0.8, 0.3, 0.01)$ and $(\mu, \mu') = (1.0, 1.1)$ and the on-site spin-orbit coupling $H_{SO} = 2\eta \sum_i \mathbf{L}_i \cdot \mathbf{S}_i$ where the sum runs over all sites of the lattice and $\eta = 0.1$. In momentum space, this yields constant coupling of longitudinal type and spin flip type $H_{SO} = \sum_{ab,ss'} \tilde{t}^{ss'}_{ab} c^\dagger_{a,s}(\mathbf{p}) c_{b,s'}(\mathbf{p})$ with the nonzero elements $\tilde{t}^{ss}_{12} = is\eta$, $\tilde{t}^{ss}_{21} = -is\eta$, $\tilde{t}^{s-s}_{13} = -i\eta$, $\tilde{t}^{s-s}_{31} = is\eta$, $\tilde{t}^{s-s}_{23} = -s\eta$, $\tilde{t}^{s-s}_{32} = -s\eta$. The Fermi surface of this model together with the density of states is shown in Fig. S2. Note that the Fermi energy is $E_F \approx 1.34t$ where the NN hopping $t$ is expected to be of the order of 100 meV[2].

For the calculation in real space with system size of $(N_x, N_y)$ lattice points in $x$ and $y$ direction, the hopping elements and contributions from the spin-orbit coupling are set up in sparse matrices to be used to calculate the eigenvalues, or the topological invariant, see below.

The order parameter for Sr$_2$RuO$_4$ is considered to be of helical-$p$ wave type with the same parametrization of the higher harmonics of the pairing components as in Ref. 1 which we state here again for convenience. The order parameters in orbital space are given by the following ansatz

$$\mathbf{d}_h^a(\mathbf{p}) = -i\Delta_t \sum_{j=1,2,3} \left( \Delta_{x,j}^a \ g_{x,j}(\mathbf{p})\mathbf{e}_y + \Delta_{y,j}^a \ g_{y,j}(\mathbf{p})\mathbf{e}_x \right) \tag{S6}$$

$$g_{x,1}(\mathbf{p}) = \sin(p_x) \tag{S7}$$
$$g_{x,2}(\mathbf{p}) = \sin(p_x)\cos(p_y) \tag{S8}$$
$$g_{x,3}(\mathbf{p}) = \sin(3p_x) \tag{S9}$$

where $a = xz, yz, xy$ is the orbital index, $g_{y,j}(p_x, p_y) = g_{x,j}(p_y, p_x)$ and $\Delta_{y,j}^{zy} = \Delta_{x,j}^{zx}$; $\Delta_{y,j}^{zx} = \Delta_{x,j}^{zy} = 0$; $\Delta_{x,j}^{xy} = \Delta_{y,j}^{xy} \ \forall \ j$ with the following parameters: $(\Delta_{x,1}^{zx}, \Delta_{x,2}^{zx}, \Delta_{x,3}^{zx}) = (0, 0.2, 1.0)$ and $(\Delta_{x,1}^{xy}, \Delta_{x,2}^{xy}, \Delta_{x,3}^{xy}) = (0.18, 0.15, -0.3)$. The magnitude of the order parameter is chosen to be small relative to the overall bandwidth of the (unrenormalized) electronic structure, $\Delta_t = 0.1$ such that the density of states in the normal state within this energy scale essentially flat, see Fig. S2(b,c). Still the DOS in the superconducting state shows the largely anisotropic order parameter with small energy gap and coherence peaks as shown in Fig. S2(c).

### Supplementary Note S2. EFFECTIVE HAMILTONIAN FOR THE SINGLE-BAND MODEL WITHOUT ZEEMAN FIELD

In the next sections, we study the effective Hamiltonian for the impurity chains along $x$ direction, with the impurity positions are parametrized as $\mathbf{r}_n = (na, 0)$ ($n \in \mathbb{Z}$), in chiral and helical $p$-wave superconductors

$$H_{\text{eff}}(k_x) = \hat{U}^{-1}\tilde{G}^{-1}\left[G_I(k_x)\hat{U} - 1\right], \tag{S10}$$

where

$$\hat{U} = V_{\text{imp}}\tau_z\sigma_0, \ \tilde{G} = \frac{1}{\Omega_{BZ}}\int_{BZ} d^2p \ [H_{\text{BdG}}(\mathbf{p}) + H_Z]^{-2}, \tag{S11}$$

and

$$G_I(k_x) = \sum_n G(0, \mathbf{r}_n)e^{-iank_x}, \ G(E, \mathbf{r}) = \frac{1}{\Omega_{\text{BZ}}}\int_{\text{BZ}} d^2p \, G(E, \mathbf{p})e^{i\mathbf{p}\cdot\mathbf{r}}, \ G(E, \mathbf{p}) = [E - H_{\text{BdG}}(\mathbf{p}) - H_Z]^{-1}. \tag{S12}$$

We start by studying the Hamiltonian in absence of a Zeeman field $\mathbf{h} = \mathbf{0}$. In this case, we obtain for both chiral and helical $p$-wave superconductors

$$[H_{\text{BdG}}(\mathbf{p})]^2 = [\xi^2(\mathbf{p}) + |\Delta(\mathbf{p})|^2]\tau_0\sigma_0, \ |\Delta(\mathbf{p})|^2 = |\Delta_t|^2(\sin^2 p_x + \sin^2 p_y) \tag{S13}$$

so that

$$\tilde{G} = \mathcal{A}\tau_0\sigma_0, \ \mathcal{A} = \frac{1}{(2\pi)^2}\int_{BZ} d^2p \frac{1}{\xi^2(\mathbf{p}) + |\Delta(\mathbf{p})|^2}, \tag{S14}$$

and

$$H_{\text{eff}}(k_x) = -\frac{1}{\mathcal{A}V_{\text{imp}}}\tau_z\sigma_0 + \frac{1}{\mathcal{A}}\tau_z\sigma_0 G_I(k_x)\tau_z\sigma_0. \tag{S15}$$

#### 1. Single impurity

In the case of single impurity the effective Hamiltonian is

$$H_{\text{eff}} = -\frac{1}{\mathcal{A}V_{\text{imp}}}\tau_z\sigma_0 + \frac{1}{\mathcal{A}}\tau_z\sigma_0 G(E = 0, \mathbf{r} = 0)\tau_z\sigma_0. \tag{S16}$$

Now

$$G(E = 0, \mathbf{r} = 0) = -\frac{1}{(2\pi)^2}\int_{BZ} d^2p \frac{1}{\xi^2(\mathbf{p}) + |\Delta(\mathbf{p})|^2}H_{\text{BdG}}(\mathbf{p}). \tag{S17}$$

By noticing that $\Delta(-\mathbf{p}) = -\Delta(\mathbf{p})$, we see that the momentum integral of $\Delta(\mathbf{p})/(\xi^2(\mathbf{p}) + |\Delta(\mathbf{p})|^2)$ vanishes, and therefore

$$G(E=0, \mathbf{r}=0) = -\mathcal{B}\tau_z\sigma_0, \quad \mathcal{B} = \frac{1}{(2\pi)^2}\int_{BZ} d^2p \frac{\xi(\mathbf{p})}{\xi^2(\mathbf{p}) + |\Delta(\mathbf{p})|^2}. \tag{S18}$$

Since only the square of the order parameter enters, the single impurity Hamiltonian for both chiral and helical p-wave superconductor is

$$H_{\text{eff}} = -\epsilon_0 \tau_z \sigma_0, \quad \epsilon_0 = \frac{1 + \mathcal{B}V_{\text{imp}}}{\mathcal{A}V_{\text{imp}}}. \tag{S19}$$

### 2. Impurity chain in a chiral p-wave superconductor

In the case of chiral p-wave superconductor without field we can write the BdG Hamiltonian as

$$H_{\text{BdG}}(\mathbf{p}) = \xi(\mathbf{p})\tau_z\sigma_0 + \Delta_t \sin p_x \tau_x \sigma_x - \Delta_t \sin p_y \tau_y \sigma_x \tag{S20}$$

and the bulk Green function at zero energy is

$$G(E=0, \mathbf{p}) = -\frac{1}{\xi^2(\mathbf{p}) + |\Delta(\mathbf{p})|^2} H_{\text{BdG}}(\mathbf{p}). \tag{S21}$$

Therefore,

$$G(E=0, \mathbf{r}_n) = -\tau_z\sigma_0 \frac{1}{(2\pi)^2}\int d^2p \frac{\xi(\mathbf{p})}{\xi^2(\mathbf{p}) + |\Delta(\mathbf{p})|^2} e^{ip_x na} - \tau_x\sigma_x \frac{1}{(2\pi)^2}\int d^2p \frac{\Delta_t \sin p_x}{\xi^2(\mathbf{p}) + |\Delta(\mathbf{p})|^2} e^{ip_x na}. \tag{S22}$$

Thus, we obtain

$$H_{\text{eff}}(k_x) = \xi_{\text{eff}}(k_x)\tau_z\sigma_0 + \Delta_{\text{eff}}(k_x)\tau_x\sigma_x, \tag{S23}$$

with

$$\xi_{\text{eff}}(k_x) = -\epsilon_0 + \sum_{n\neq 0} h_n e^{ik_x n}, \quad h_n = -\frac{1}{\mathcal{A}}\frac{1}{(2\pi)^2}\int d^2p \frac{\xi(\mathbf{p})}{\xi^2(\mathbf{p}) + |\Delta(\mathbf{p})|^2} e^{-ip_x na}, \tag{S24}$$

and

$$\Delta_{\text{eff}}(k_x) = \sum_{n\neq 0} \Delta_n e^{ik_x n}, \quad \Delta_n = \frac{1}{\mathcal{A}}\frac{1}{(2\pi)^2}\int d^2p \frac{\Delta_t \sin p_x}{\xi^2(\mathbf{p}) + |\Delta(\mathbf{p})|^2} e^{-ip_x na}. \tag{S25}$$

By reordering the Nambu basis from $\psi_{\mathbf{k}}^\dagger = (c_{\mathbf{k}\uparrow}^\dagger, c_{\mathbf{k}\downarrow}^\dagger, c_{-\mathbf{k}\uparrow}, c_{-\mathbf{k}\downarrow})$ to $\tilde{\psi}_{\mathbf{k}}^\dagger = (c_{\mathbf{k}\uparrow}^\dagger, c_{-\mathbf{k}\downarrow}, c_{\mathbf{k}\downarrow}^\dagger, c_{-\mathbf{k}\uparrow})$, we can turn the Hamiltonian into a block-diagonal form

$$\tilde{H}_{\text{eff}}(k_x) = \begin{pmatrix} \tilde{H}_0(k_x) & 0 \\ 0 & \tilde{H}_0(k_x) \end{pmatrix} = \tilde{\sigma}_0 \tilde{H}_0(k_x), \quad \tilde{H}_0(k_x) = \begin{pmatrix} \xi_{\text{eff}}(k_x) & \Delta_{\text{eff}}(k_x) \\ \Delta_{\text{eff}}(k_x) & -\xi_{\text{eff}}(k_x) \end{pmatrix} = \xi_{\text{eff}}(k_x)\tilde{\tau}_z + \Delta_{\text{eff}}(k_x)\tilde{\tau}_x. \tag{S26}$$

Here $\tilde{\tau}_i$ and $\tilde{\sigma}_i$ are Pauli matrices in the new basis which still correspond to particle-hole and spin degrees of freedom. With the help of these matrices we can write $\tilde{H}_{\text{eff}}(k_x) = \xi_{\text{eff}}(k_x)\tilde{\sigma}_0\tilde{\tau}_z + \Delta_{\text{eff}}(k_x)\tilde{\sigma}_0\tilde{\tau}_x$.

Although the Hamiltonian is block-diagonal, these blocks are not independent degrees of freedom because the particle-hole symmetry connects these. Let us remind the reader that particle-hole symmetry is in the original basis

$$\tau_x\sigma_0 H_{\text{eff}}^T(-k_x) \tau_x\sigma_0 = -H_{\text{eff}}(k_x) \tag{S27}$$

and it now reads in the new basis as

$$\tilde{\sigma}_x\tilde{\tau}_x \tilde{H}_{\text{eff}}^T(-k_x) \tilde{\sigma}_x\tilde{\tau}_x = -\tilde{H}_{\text{eff}}(k_x). \tag{S28}$$

The fact that the particle-hole symmetry is $\tilde{\tau}_x\tilde{\sigma}_x$ means that if there is a positive energy solution in the first block there is a negative energy solution in the second block.

### 3. Impurity chain in a helical p-wave superconductor

In the case of helical $p$-wave superconductor without Zeeman field we can write the BdG Hamiltonian as

$$H_{BdG}(\mathbf{p}) = \xi(\mathbf{p})\tau_z\sigma_0 + \Delta_t \sin p_x \tau_x \sigma_0 - \Delta_t \sin p_y \tau_y \sigma_z \tag{S29}$$

and

$$G(E=0, \mathbf{r}_n) = -\tau_z\sigma_0 \frac{1}{(2\pi)^2} \int d^2 p \frac{\xi(\mathbf{p})}{\xi^2(\mathbf{p}) + |\Delta(\mathbf{p})|^2} e^{ip_x na} - \tau_x\sigma_0 \frac{1}{(2\pi)^2} \int d^2 p \frac{\Delta_t \sin p_x}{\xi^2(\mathbf{p}) + |\Delta(\mathbf{p})|^2} e^{ip_x na}. \tag{S30}$$

Thus, we obtain the effective Hamiltonian as stated in the main text

$$H_{\text{eff}}(k_x) = \xi_{\text{eff}}(k_x)\tau_z\sigma_0 + \Delta_{\text{eff}}(k_x)\tau_x\sigma_0. \tag{S31}$$

Because of the common factor $\sigma_0$, this Hamiltonian is already formally block-diagonal without basis transformation. The important difference to the case for the chiral order parameter is that here the particle-hole symmetry still operators inside each block since it still has the form $\sigma_0\tau_x$.

## Supplementary Note S3. EFFECTIVE HAMILTONIAN IN THE PRESENCE OF THE ZEEMAN FIELD

In this section we include the Zeeman term $H_Z = \mathbf{h} \cdot \boldsymbol{\sigma}$ in the Hamiltonian and study how the magnitude and direction of the Zeeman field $\mathbf{h} = (h_x, h_y, h_z)$ influences the topological properties of the system. We assume that the $|\mathbf{h}| \ll \Delta_t$, so that the Zeeman field does not influence the superconducting order parameter [3]. The effect of the Zeeman field can be analyzed numerically using the effective Hamiltonian (S10). Here we try to give an analytically transparent expressions for the effect of Zeeman field.

The first simplification of the effective Hamiltonian $H_{\text{eff}}(k_x)$ is obtained by noticing that $\tau_z\sigma_0\tilde{G}^{-1}$ exists as a common factor in $H_{\text{eff}}(k_x)$ and therefore its exact structure could be important for the topology only if the Zeeman field would cause a gap closing in the bulk Hamiltonian $H_{\text{BdG}}(\mathbf{p})$. We will consider only weak Zeeman fields which do not cause gap closings in the bulk. Therefore, without modifying the topology of the effective Hamiltonian $H_{\text{eff}}(k_x)$ for the impurity chain, we can evaluate $\tilde{G}$ in the absence of the Zeeman field.

Although we have not managed to evaluate $G_I(k_x)$ analytically in the case of the general direction of the Zeeman field, we have obtained approximate expressions for the effective Hamiltonian. We express these results by utilizing the matrix obtained for one block of the Hamiltonian in the previous section

$$\tilde{H}_0(t, \mu, \Delta_t, V_{\text{imp}}, a, k_x) = \begin{pmatrix} \xi_{\text{eff}}(t, \mu, \Delta_t, V_{\text{imp}}, a, k_x) & \Delta_{\text{eff}}(t, \mu, \Delta_t, V_{\text{imp}}, a, k_x) \\ \Delta_{\text{eff}}(t, \mu, \Delta_t, V_{\text{imp}}, a, k_x) & -\xi_{\text{eff}}(t, \mu, \Delta_t, V_{\text{imp}}, a, k_x) \end{pmatrix}. \tag{S32}$$

The dependence of the effective dispersion $\xi_{\text{eff}}(t, \mu, \Delta_t, V_{\text{imp}}, a, k_x)$ and pairing $\Delta_{\text{eff}}(t, \mu, \Delta_t, V_{\text{imp}}, a, k_x)$ on the parameters $t, \mu, \Delta_t, V_{\text{imp}}, a$ and $k_x$ is determined by the equations described in the previous section. Notice that although the Hamiltonian is not necessarily exactly block-diagonal in the presence of the Zeeman field, we find that it can always be expressed approximately in a block-diagonal form in the suitable basis. In the following, we directly express the results in the basis where the Hamiltonian is approximately block-diagonal.

### A. Chiral p-wave superconductor

#### 1. Zeeman field along z-direction

In the case of chiral $p$-wave superconductor and Zeeman field in $z$-direction $\mathbf{h} = (0, 0, h_z)$ the effective model for the impurity chain is

$$\tilde{H}_{\text{eff}}(k_x) = \begin{pmatrix} \tilde{H}_0(t, \mu, \Delta_t, V_{\text{imp}}, a, k_x) + h_z\tau_0 & 0 \\ 0 & \tilde{H}_0(t, \mu, \Delta_t, V_{\text{imp}}, a, k_x) - h_z\tau_0 \end{pmatrix}. \tag{S33}$$

Here, the Zeeman field trivially shifts one block upwards in energy and the other block downwards in energy. Thus, it cannot cause a transition to a topologically nontrivial state. However, the Zeeman field causes a transition from gapped phase into a gapless phase.

### 2. Zeeman field along x- or y-direction

In the case of chiral $p$-wave superconductor and Zeeman field in $x$- or $y$-direction $\mathbf{h} = (h_x, 0, 0)$ or $\mathbf{h} = (0, h_y, 0)$ the effective model for the impurity chain is

$$\tilde{H}_{\text{eff}}(k_x) = \begin{pmatrix} \tilde{H}_0(t, \mu + h_{x(y)}, \Delta_t, V_{\text{imp}}, a, k_x) & 0 \\ 0 & \tilde{H}_0(t, \mu - h_{x(y)}, \Delta_t, V_{\text{imp}}, a, k_x) \end{pmatrix}. \tag{S34}$$

The Zeeman term enters the effective Hamiltonian indirectly via renormalization of the chemical potential $\mu \to \mu \pm h_{x(y)}$ in the two blocks. This can cause topological phase transitions but typically, the impurity bound state energy, and thus the impurity bands are only weakly dependent on the chemical potential via the change of the (normal state) density of states at the Fermi level (see Fig. S7). Thus, one needs relatively strong Zeeman field to cause a transition and the topological gap stays small.

## B. Helical $p$-wave superconductor

### 1. Zeeman field along z-direction

In the case of helical $p$-wave superconductor and Zeeman field in $z$-direction $\mathbf{h} = (0, 0, h_z)$ the effective model for the impurity chain is

$$\tilde{H}_{\text{eff}}(k_x) = \begin{pmatrix} \tilde{H}_0(t, \mu + h_z, \Delta_t, V_{\text{imp}}, a, k_x) & 0 \\ 0 & \tilde{H}_0(t, \mu - h_z, \Delta_t, V_{\text{imp}}, a, k_x) \end{pmatrix}. \tag{S35}$$

Therefore the topological phase diagram is exactly the same as in the case of chiral $p$-wave superconductor with Zeeman field in $x$- or $y$-direction.

### 2. Zeeman field along x-direction

In the case of helical $p$-wave superconductor and Zeeman field in $x$-direction $\mathbf{h} = (h_x, 0, 0)$ the effective model for the impurity chain is

$$\tilde{H}_{\text{eff}}(k_x) = \begin{pmatrix} \tilde{H}_0(t, \mu, \Delta_t, V_{\text{imp}}, a, k_x) + h_{\text{eff},x}\tau_z & 0 \\ 0 & \tilde{H}_0(t, \mu, \Delta_t, V_{\text{imp}}, a, k_x) - h_{\text{eff},x}\tau_z \end{pmatrix}. \tag{S36}$$

Here we have made approximations during the derivation of the effective Hamiltonian, so that the expression only serves as a good approximation for calculation of the topological phase diagram. The Zeeman field in this case is very effective in causing topological phase transitions. It acts almost directly to the impurity states. The magnitude of the effective Zeeman field $h_{\text{eff},z}$ is renormalized from the bare value of Zeeman field $h_{\text{eff},z} \approx h_x/2$.

### 3. Zeeman field along y-direction

In the case of helical $p$-wave superconductor and Zeeman field in $y$-direction $\mathbf{h} = (0, h_y, 0)$ the effective model for the impurity chain is

$$\tilde{H}_{\text{eff}}(k_x) = \begin{pmatrix} \tilde{H}_0(t, \mu, \Delta_t, V_{\text{imp}}, a, k_x) + h_{\text{eff},y}\tau_0 & 0 \\ 0 & \tilde{H}_0(t, \mu, \Delta_t, V_{\text{imp}}, a, k_x) - h_{\text{eff},y}\tau_0 \end{pmatrix}. \tag{S37}$$

Here we have also made some approximations. After these approximations it seems that the effect Zeeman field in this case is similar as in the case of chiral $p$-wave superconductor with Zeeman field along $z$-direction. Therefore, it just shifts the blocks in different directions in energy and cannot induce a topological transition. However, it causes a transition from gapped phase into a gapless phase. The magnitude of the effective Zeeman field $h_{\text{eff},y}$ is again renormalized from the bare value of Zeeman field $h_y$.

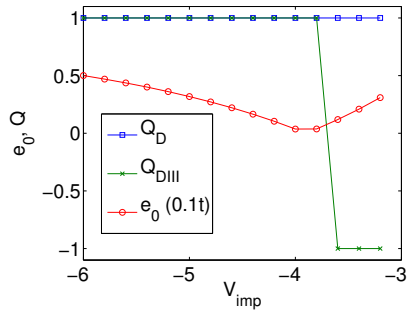

FIG. S3. Magnitude of the lowest eigenvalue $e_0$ together with the class D and class DIII invariants showing that the class DIII invariant changes when the eigenvalue hits zero. The calculation is done for impurity spacing $a = 15$ using a supercell of size $N_x = 15$ sites along the chain and $N_y = 35$ sites perpendicular to the chain.

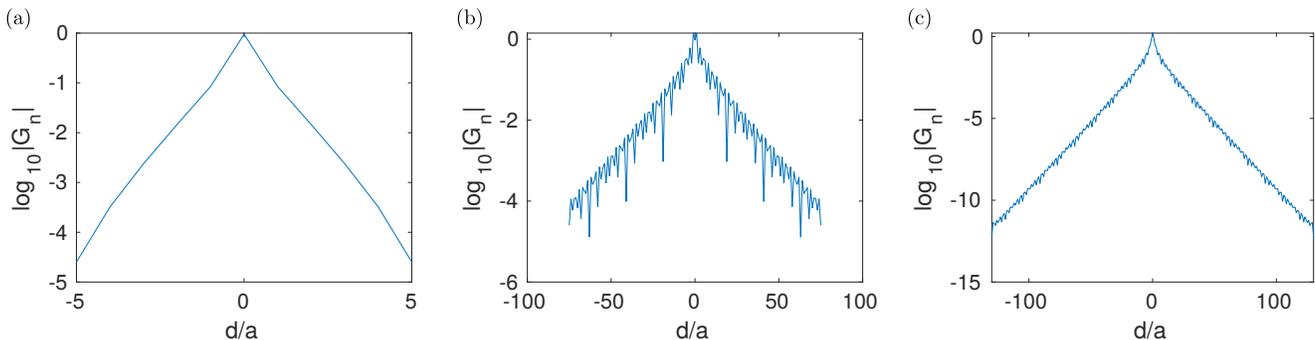

FIG. S4. Norm of the expansion coefficients of the quasiparticle propagator showing exponential decay with distance for the different models: (a) Single-band model with chiral $p$-wave order parameter. (b) Single-band model with helical $p$-wave order parameter, which is strongly anisotropic so that the gap is suppressed by a factor of 10 along $p_y$. (c) Multiband model for $Sr_2RuO_4$ with helical $p$-wave order parameter, which is strongly anisotropic so that there is a gap minima along the 45 degree direction in $(p_x, p_y)$-plane.

## Supplementary Note S4.  TOPOLOGICAL PHASE TRANSITION WITHOUT ZEEMAN FIELD

In the case of helical $p$-wave superconductivity without the Zeeman field the system satisfies a time-reversal symmetry. Thus, it belongs to class DIII in the Altland Zirnbauer classification scheme allowing for a possibility of a topological phase transition as a function of the impurity strength. In the topologically nontrivial phase, there exists two degenerate Majorana zero modes at each end of the impurity chain. In the case of chiral $p$-wave superconductivity without Zeeman field the system supports a spin-rotation symmetry that allows to block-diagonalize the Hamiltonian. Thus also in this case the system can support a topologically nontrivial phase with two degenerate Majorana zero modes appearing at each end of the chain. (To be more precise the effective Hamiltonian in both cases also supports a chiral symmetry allowing infinite number of topologically distinct phases to appear, but the Majorana end modes always appear in pairs in the absence of the Zeeman field.)

Because the Majorana end modes appear in pairs they are not so useful for topological quantum computing. However, we can check the numerical implementation by calculating the corresponding DIII invariant as described in Ref. 4 which is based on the calculation of the product of Pfaffians at time reversal invariant momenta. We therefore tune the impurity band through the chemical potential by varying the impurity potential $V_{imp}$ from below $V_{imp}^*$ to above that value. Indeed, the topological invariant as calculated numerically from the Pfaffian changes sign when the energy of the bound state $e_0$ hits zero. When looking at the energy bands as function of $k_x$ one can also observe that such an impurity band is pushed through zero in this case. Due to time-reversal symmetry, the eigenvalues still come in pairs, therefore the class D invariant (as described in the main text) stays at $+1$ as expected, see Fig. S3.

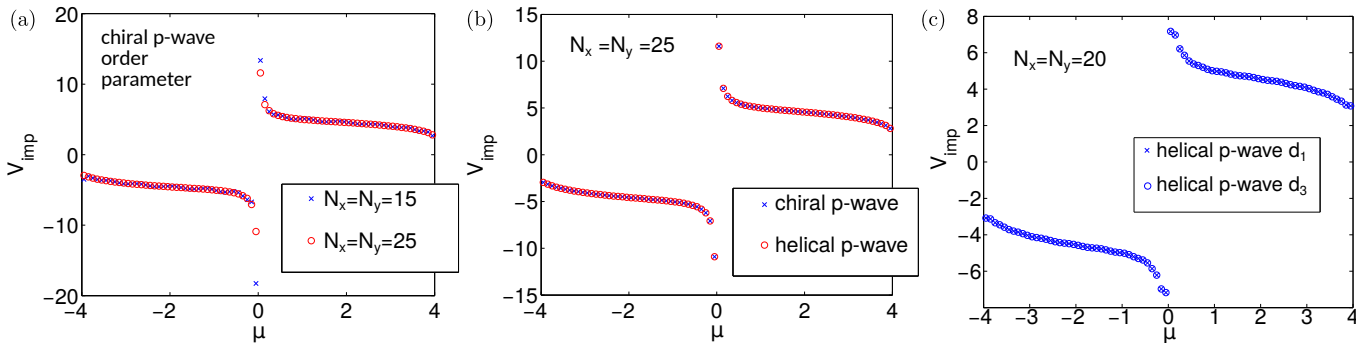

FIG. S5. Results for the impurity potential $V^*_{\text{imp}}$ that yields a bound state at zero energy as a function of the chemical potential. The order parameter is set to $\Delta_t = 0.5$. (a) Already at a system size of $N_x = N_y = 25$, no finite size effects are visible, while $N_x = N_y = 15$ show some oscillations from discrete energy levels. (b) The bound state energy is exactly identical for chiral-$p$ wave order parameter, Eq. (S5), and helical p-wave order parameter, Eq. (S4), as expected from the analytical result, Eq. (S19). (c) The same is true when comparing the results for the different basis functions, Eq. (S2).

## Supplementary Note S5.    NUMERICAL RESULTS FOR THE ENERGY OF THE IMPURITY BOUND STATE

The effective Hamiltonian Eq. (S10) describes the topological phase transitions (energy gap closings) exactly, provided that the Brillouin zone integrals are evaluated with sufficient accuracy and all the longer range hoppings and pairing amplitudes, given for example in Eqs. (S24) and (S25), are included. These longer range terms are expected to decay exponentially with distance because the bulk Hamiltonian is fully gapped. To verify this, we have examined the norm of $G(0, \mathbf{r}_n)$ as function of distance $d = |\mathbf{r}_n|$ for the various models. The exponential decay is demonstrated in Fig. S4 for all models considered indicating that the errors are exponentially small if the expansion is truncated at finite $\mathbf{r}_n$.

We have also numerically studied the tight-binding models using finite size supercells with the impurity in the center. This approach leads to finite size effects in the energy of the order of the bandwith divided by the number of quantum states. To estimate the required system sizes, we perform a check of finite size effects by varying the system size and calculating the impurity potential $V^*_{\text{imp}}$ where the bound state robustly crosses the zero energy due to a change of a topological invariant as described in detail in Ref. 5 (see Fig. S5). Plotting this quantity as function of the chemical potential $\mu$ one can easily estimate the effects of the energy spacing as small oscillations. These can be seen in Fig. S5(a) for a system size of 15x15 elementary cells (single band model with isotropic order parameter), while for 25x25 elementary cells these effects are not present any more. According to Eq. (S19) the bound state energy is the same for chiral and helical $p$-wave order parameters because only the absolute magnitude of the gap enters the calculation, and this is verified also numerically in Fig. S5(b-c). Note further that the results of $V^*_{\text{imp}}(\mu)$ are very flat (except close to the van Hove singularity at $\mu = 0$). As explained above, this property makes it very difficult to control the topological phase transition by the use of a Zeeman field in $z$ direction in the case of helical $p$-wave order parameter and an in-plane field in the case of chiral $p$-wave order parameter.

To compare our analytical result [Eq. (S19)] with the numerical implementation, we calculate the bound state energy at a fixed filling of $n = 0.25$ as a function of the impurity potential $V_{\text{imp}}$ and show the result in Fig. S6. The zero-energy crossing is captured exactly, while there are small deviations at non-zero energies arising from the expansion of the Green function in powers of the energy. Finite size effects of the numerical implementation are clearly seen when plotting the impurity bound state energy as function of chemical potential (see Fig. S7). Note again that the bound state energy depends only weakly on the chemical potential.

## Supplementary Note S6.    NUMERICAL RESULTS FOR IMPURITY BANDS IN THE PRESENCE OF ZEEMAN FIELD

In Fig. S8 we show the impurity bands for chiral $p$-wave superconductor in the presence of the Zeeman field. If Zeeman field is applied in-plane it only weakly breaks the degeneracy of the impurity bands [see Fig. S8(a),(b)] as expected from Eq. (S34). Therefore, a strong field is required to induce a topological phase transition. If Zeeman field is applied along $z$-direction the degeneracy of the impurity bands is broken strongly so that one band is shifted up and the other down in energy [Fig. S8(c)] such that the bands crosss yielding a gapless system as expected from

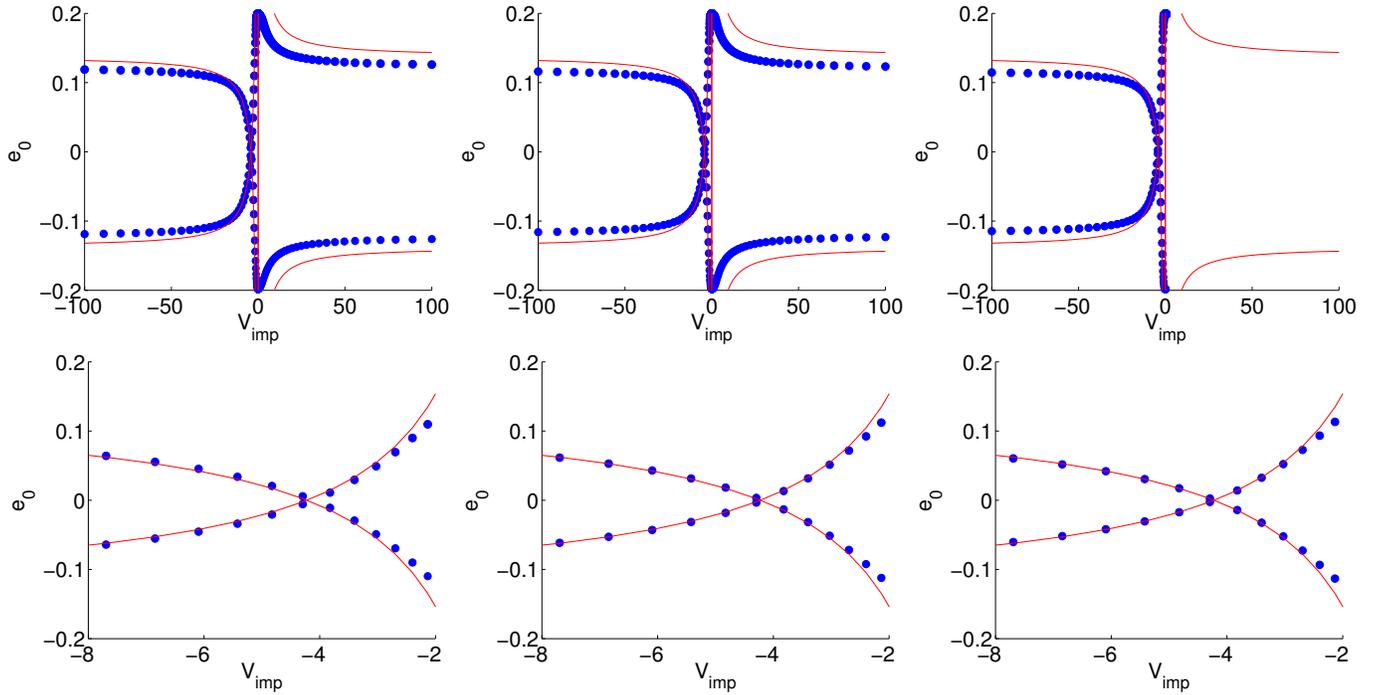

FIG. S6. Bound state energy as a function of the impurity potential. It crosses the zero-energy at $V_{\text{imp}} = V_{\text{imp}}^*$. Bottom row shows a zoom-in on the the crossing. Size of the real space lattice from left to right: $N_x = N_y = 30, 40, 60$. Red curves are evaluated using the effective Hamiltonian (S19).

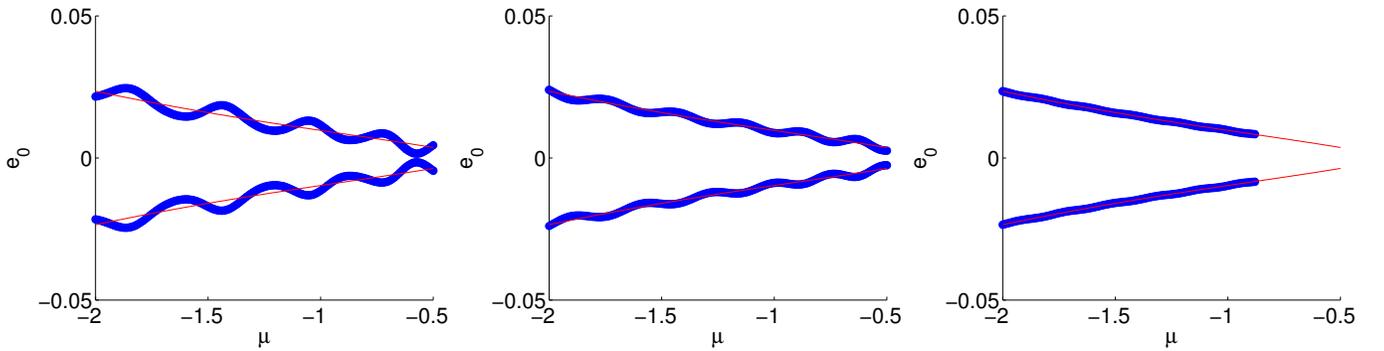

FIG. S7. Bound state energy as a function of the chemical potential at fixed impurity potential $V_{\text{imp}} = -4.7416$. Red curves are evaluated using the the effective Hamiltonian (S19). Left to right: $N_x = N_y = 30, 40, 60$.

Eq. (S33).

In Fig. S9 we show the impurity bands for helical $p$-wave superconductor in the presence of the Zeeman field. If Zeeman field is applied in $x$-direction the degeneracy of the impurity bands is strongly broken, but apart from the topological phase transition point the system remains gapped [Fig. S9(a)] as expected from Eq. (S36). If Zeeman field is applied along the $y$-direction the bands are just shifted in energy and the system becomes gapless [Fig. S9(b)] as expected from Eq. (S37). Finally, the field in $z$ direction affects the bands only very weakly [Fig. S9(c)] so that a strong field is required to induce a topological phase transition as expected from Eq. (S35).

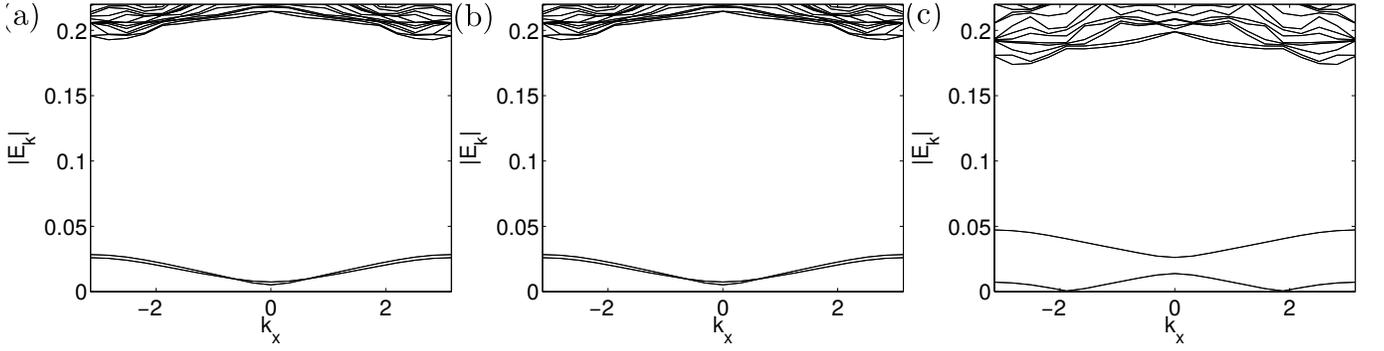

FIG. S8. Impurity bands for the chiral $p$-wave superconductor with $V_{\mathrm{imp}} = -4.8$, $a = N_x = 15$, $N_y = 35$ and Zeeman field $|\mathbf{h}| = 0.02$ applied in (a) $x$, (b) $y$ and (c) $z$ direction.

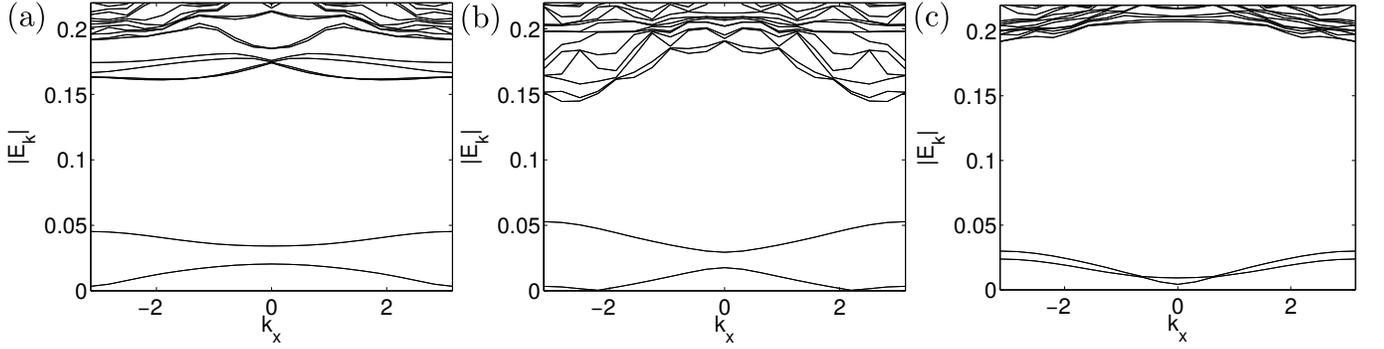

FIG. S9. Impurity bands for the helical $p$-wave superconductor with $V_{\mathrm{imp}} = -4.8$, $a = N_x = 15$, $N_y = 35$ and Zeeman field $|\mathbf{h}| = 0.05$ applied in (a) $x$, (b) $y$ and (c) $z$ direction.

### Supplementary Note S7. PHASE DIAGRAMS

With the discussion of the effects of the external Zeeman field in various directions, we can also understand the general properties of the phase diagram obtained by calculating the topological invariant

$$Q = \prod_{k_x \in \mathrm{TRIM}} Q(k_x). \tag{S38}$$

We have studied the topological phase diagram both by using the full tight-binding Hamiltonian and the effective Hamiltonian (Figs. S10, S11, S12 and S13).

Figs. S10 and S11 show $Q$ for the chiral $p$-wave superconductor as a function $|\mathbf{h}|$ and $V_{\mathrm{imp}}$ as obtained from the effective Hamiltonian and the full tight-binding Hamiltonian, respectively. The results are in excellent agreement with each other. The in-plane fields can cause a topological phase-transition to a topologically nontrivial phase with $Q = -1$ [Figs. S10(a),(b) and S11(a),(b)], but it requires large fields and the magnitude of the topological gap remains relatively small [Fig. S14(d),(e)]. Zeeman field applied in the $z$-direction can make $Q = -1$ [Figs. S10(c) and S11(c)] but in this case the system is gapless [Fig. S14(f)].

Figs. S12 and S13 show similar phase diagrams for the helical $p$-wave superconductor. The Zeeman field in $x$-direction is very effective in causing a transition to a topologically nontrivial phase with $Q = -1$ [Figs. S12(a) and S13(a)] leading to a large topological energy gap [Fig. S14(a)]. Zeeman field applied in the $y$-direction can make $Q = -1$ [Figs. S12(b) and S13(b)] but in this case the system is gapless [Fig. S14(b)]. Finally, the Zeeman field applied in $z$-direction can cause a transition to a topologically nontrivial phase with $Q = -1$ [Figs. S12(c) and S13(c)], but it requires large fields and the magnitude of the topological gap remains relatively small [Fig. S14(c)].

The dependence of the topological phase diagram on the direction of the Zeeman field [Fig. S15] is strikingly different in the cases of helical and chiral $p$-wave superconductors. Thus, it can be used as a diagnostic tool to determine the order parameter symmetry of triplet superconductors.

In Fig. S16 we show that the topologically nontrivial phase can be reached also by placing the impurities at the step edge appearing on the surface of the system.

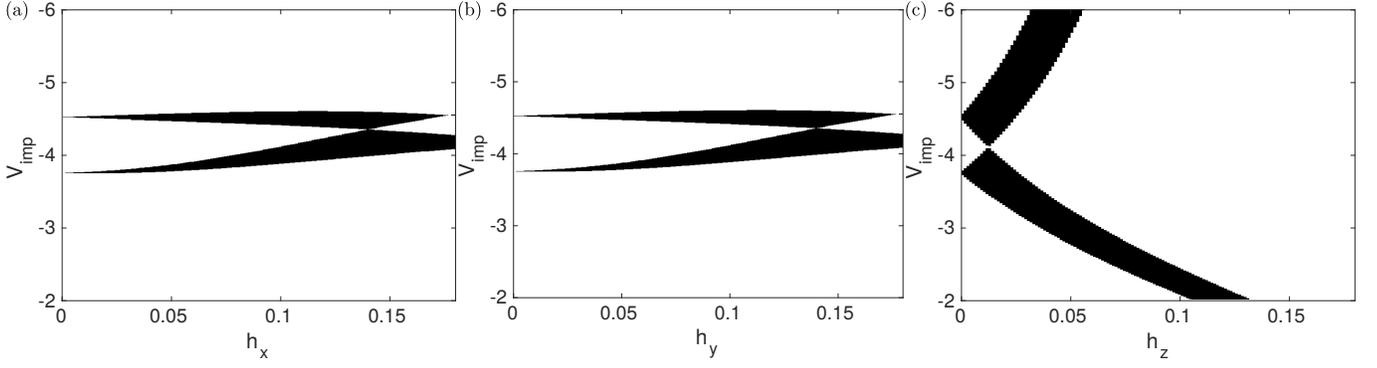

FIG. S10. Phase diagram (white $Q = +1$, black $Q = -1$) for the chiral $p$-superconductor as obtained from the effective Hamiltonian with $a = 15$ and $\Delta_t = 0.2$. Note that for the field in $z$ direction, the system remains gapless, i.e. the zero energy states are not localized at the ends of the impurity chain, but are bulk states instead.

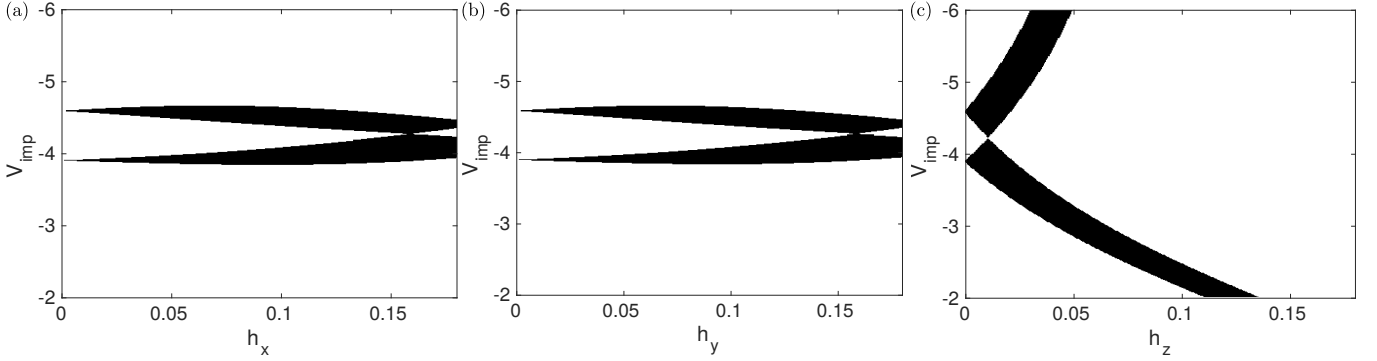

FIG. S11. Phase diagram for the chiral $p$-wave superconductor as obtained from the tight-binding model with $a = N_x = 15$, $N_y = 35$ and $\Delta_t = 0.2$.

## Supplementary Note S8.  CURVED CHAINS: MANIPULATION OF PHASE BOUNDARIES

To illustrate the tuneability of the phase boundaries, we consider a Y-junction geometry of two curved chains with different lattice constants (see Fig. S17), where Majorana zero modes appear at the interfaces of topologically trivial (black) and nontrivial (red) regimes which can be shifted by controlling the magnetic field direction. The local phase diagrams on the symmetrically placed example points $P_1$ and $P_2$ are shown in Fig. S17(c) and (d), respectively, to demonstrate that for a field with azimuthal angle $\phi = 0$ and polar angle of $\theta = 0.7\pi$, point $P_1$ is in the topologial phase and $P_2$ on the chain with the different lattice constant is in the trivial phase. Note that the phase diagrams

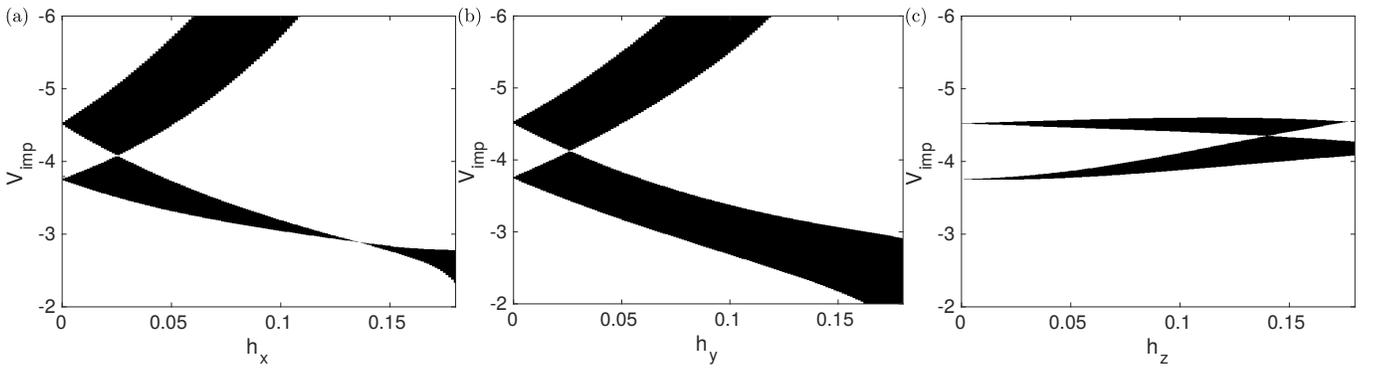

FIG. S12. Phase diagram for the helical $p$-wave superconductor obtained from the effective Hamiltonian with $a = 15$ and $\Delta_t = 0.2$.


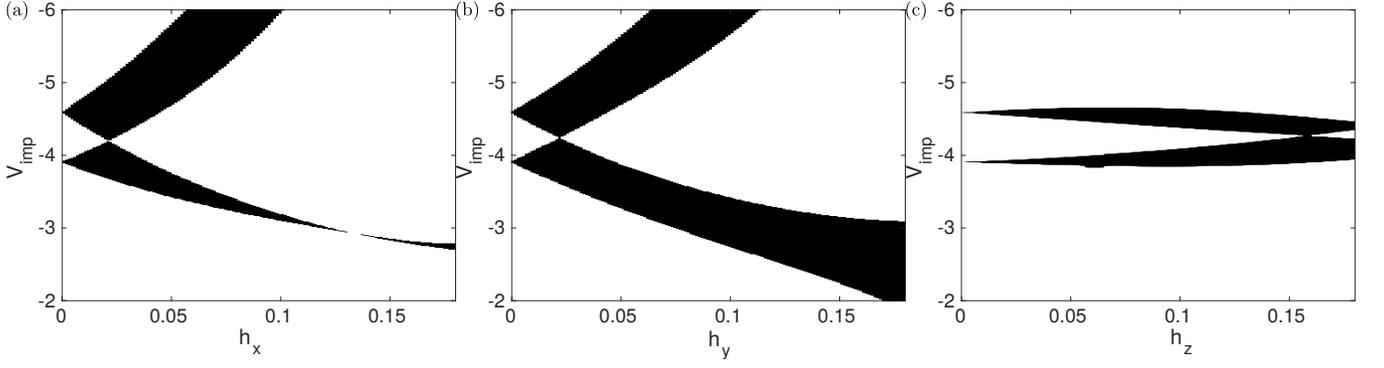

FIG. S13. Phase diagram for the helical $p$-wave superconductor as obtained from the tight-binding model with $a = N_x = 15$, $N_y = 35$ and $\Delta_t = 0.2$.

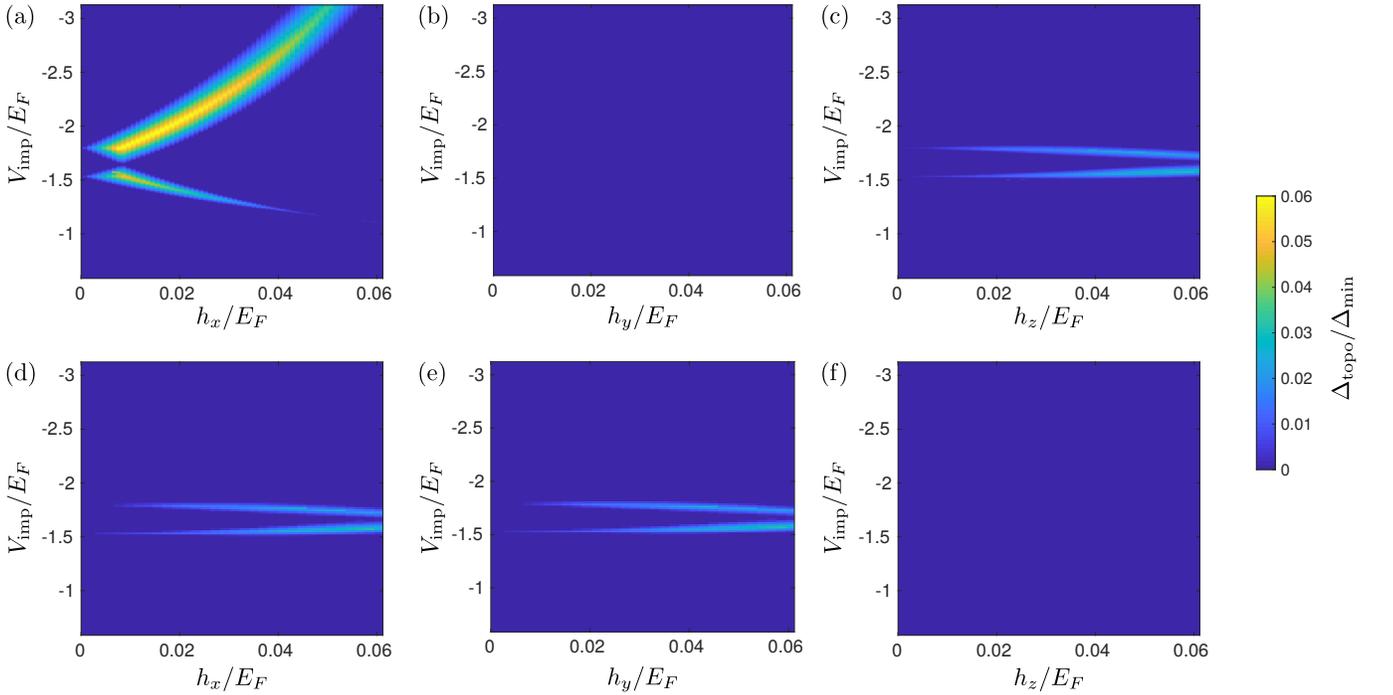

FIG. S14. (a-c) The topological energy gaps for the helical $p$-wave superconductor. (d-f) The same for the chiral $p$-wave order parameter. The parameters are $a = N_x = 15$, $N_y = 35$ and $\Delta_t = 0.2$.

(e,f) are in principle shifted by $\pi/2$ and additionally the topological phase is larger in (e) due to the different choice of the lattice constant. By rotating the Zeeman field such that it follows the trajectory drawn as red path in panels (e) and (f), one can in principle achieve an interesting movement of the phase boundaries. First the phase boundary labeled 2 moves completely down and number 3 joins on the junction (panel (b)). Rotating the field further, one can tune the system such that a segment on the left is in the trivial state and boundary 3 moves to the left. Now, one can move boundary 2 to the right hand of the Y-junction before boundary 3 moves back to the junction. In summary, we have executed a full circle in the parameter space, i.e. the magnetic field at the beginning has the same value as at the end. At this point we note that the trajectories in the phase diagrams in Fig.S17(e,f) are very close to the phase boundaries (dashed white line), i.e. the topological gap turns out to be very small such that the MZMs are less localized and tend to overlap. The experimental realization of this protocol to exchange MZMs is therefore more than challenging. More sophisticated trajectories for the external field $\mathbf{h}$ might help a bit; we have not attempted to optimize this procedure, but instead propose to use local magnets to overcome this difficulty.

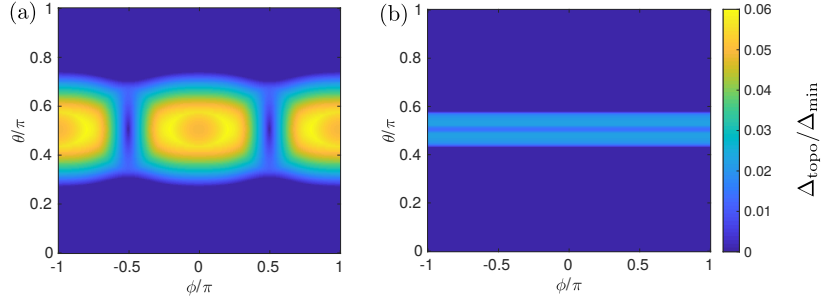

FIG. S15. Phase diagram as function of the direction of the field for (a) helical $p$-wave superconductor and (b) chiral $p$-wave superconductor. The parameters are $a = N_x = 15$, $N_y = 35$, $\Delta_t = 0.2$. $|h| = 0.1$ and $V_{\text{imp}} = 3.9$.

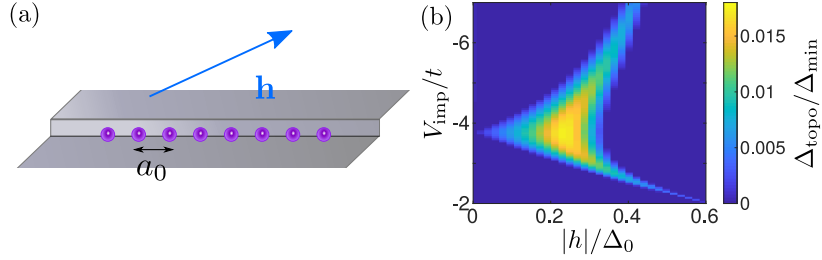

FIG. S16. Impurity chain on step edge. (a) Placing adatoms onto the surface might be easier if done along step edges. (b) Calculated topological gap for such an assembly of adatoms on step edges. We have assumed helical $p$-wave order parameter and the topological phase can be reached by application of a Zeeman field, but in this case a component perpendicular to the chain is needed. Here, $\mathbf{h} \approx |h|(0.3, 0.95, 0)$. The other parameters the same as in the main text, i.e $\Delta_t/E_F \approx 0.08$, $a \approx 2.5\xi$, while we have used the fully numerical approach with $N_y = 75$.

### Supplementary Note S9. FUSION OF MAJORANA ZERO MODES

As discussed in the main text, on curved impurity chains, two pairs of Majorana zero modes (MZMs) are generated upon increasing the Zeeman field along the chain. The corresponding zero modes can be described by the anticommuting operators $\gamma_i$, $i = 1, 2, 3, 4$ with $\{\gamma_i, \gamma_j\} = 2\delta_{ij}$. These are related to ordinary fermionic operators $c$ and $d$ with $\{c, c^\dagger\} = \{d, d^\dagger\} = 1$ with

$$c = \frac{1}{2}(\gamma_1 + i\gamma_2), \quad d = \frac{1}{2}(\gamma_3 + i\gamma_4). \tag{S39}$$

The occupation operators of these fermions are given by

$$n_c = c^\dagger c = \frac{1 + i\gamma_1\gamma_2}{2}, \quad n_d = d^\dagger d = \frac{1 + i\gamma_3\gamma_4}{2}. \tag{S40}$$

and the many-particle ground states can be defined as ($c|00\rangle = 0$, $d|00\rangle = 0$)

$$|00\rangle, \ c^\dagger|00\rangle = |10\rangle, \ d^\dagger|00\rangle = |01\rangle, \ c^\dagger d^\dagger|00\rangle = |11\rangle. \tag{S41}$$

The back transformation reads

$$\gamma_1 = c + c^\dagger, \gamma_2 = i(c^\dagger - c), \gamma_3 = d + d^\dagger, \ \gamma_4 = i(d^\dagger - d). \tag{S42}$$

In order to describe the state after the fusion, we introduce another set of fermionic operators as

$$e = \frac{1}{2}(\gamma_1 + i\gamma_4), \ f = \frac{1}{2}(\gamma_2 + i\gamma_3). \tag{S43}$$

The degenerate many-particle ground states can be written using these fermion operators as ($e|00\rangle_{e,f} = 0$, $f|00\rangle_{ef} = 0$)

$$|00\rangle_{e,f}, \ e^\dagger|00\rangle_{e,f} = |10\rangle_{e,f}, \ f^\dagger|00\rangle_{e,f} = |01\rangle_{e,f}, \ e^\dagger f^\dagger|00\rangle_{e,f} = |11\rangle_{e,f}. \tag{S44}$$

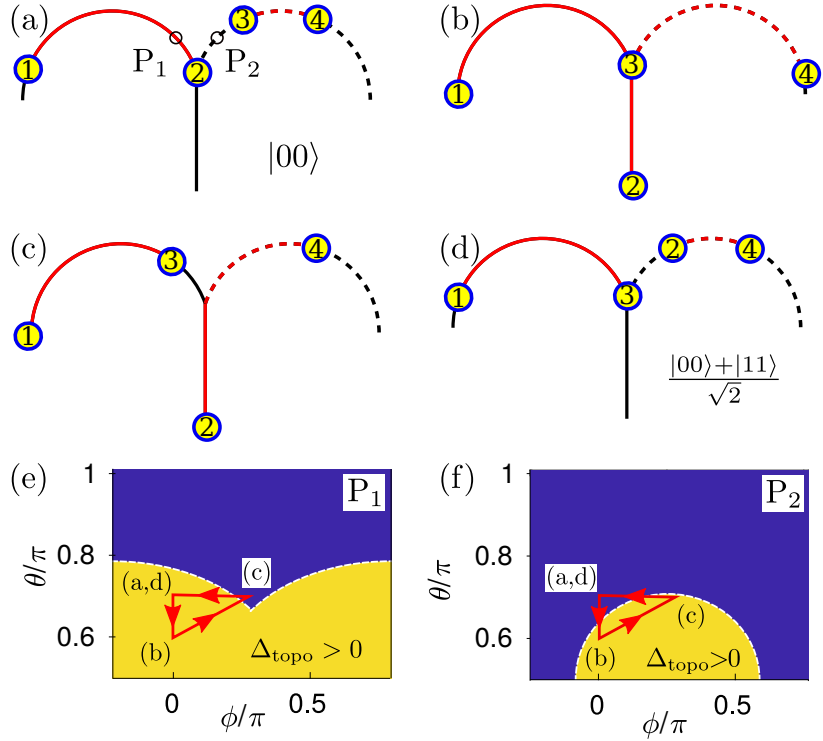

FIG. S17. Nontrival movement of phase boundaries. A Y junction can be realized by joining two curved impurity chains with different lattice constants (full line and dashed line). MZMs, labelled 1-4 are at the boundaries between topological regions (red) and trivial regions (black). (a-f) Changing the direction of the Zeeman field moves the boundaries when the direction of the magnetic field follows a closed loop in the $(\phi, \theta)$ plane shown in (e) and (f). (e),(f) Topological phase diagrams at points $P_1$ and $P_2$ (open black circles in panel (a)). The differences in the two phase diagrams are due to the different tangential directions of the chains at these points and the different lattice constants. The magnitude of the topological gap (not shown) remains small along the trajectory, thus MZMs will almost certainly overlap. In the calculation of the topological phase diagrams we have assumed lattice constants (e) $a_0 \approx 7\xi$ and (f) $a_0 \approx 6.2\xi$. Other parameters are identical to the ones in the main text, $\Delta_0/E_F \approx 0.08$.

In order to find the basis transformation between the states , we

$$e^\dagger = \frac{1}{2}(c + c^\dagger - d + d^\dagger), \ f^\dagger = \frac{i}{2}(-c + c^\dagger - d - d^\dagger). \tag{S45}$$

We observe that the unitary transformation between the two bases does not change the total parity. Therefore, using the equations (S45) we find that the basis transfomation between states (S41) and (S44) is given by

$$|00\rangle_{e,f} = \frac{1}{\sqrt{2}}(|00\rangle + |11\rangle),$$
$$|10\rangle_{e,f} = \frac{1}{\sqrt{2}}(|10\rangle + |01\rangle),$$
$$|01\rangle_{e,f} = \frac{i}{\sqrt{2}}(|10\rangle - |01\rangle),$$
$$|11\rangle_{e,f} = \frac{i}{\sqrt{2}}(|00\rangle - |11\rangle). \tag{S46}$$

From these expressions it is clear that when the MZMs 2 and 3 are fused corresponding to projective measurement of $f^\dagger f$, the measurement outcomes 0 and 1 have equal probabilities. After the measurement the system has equal probabilities to be in the two different eigenstates of $e^\dagger e$.

As discussed in the main text the MZMs can be moved by varying the direction and the magnitude of the magnetic field. Moreover, the parity of the Majoranas can be measured as soon as the MZMs 2 and 3 hybridize and acquire

a charge[6]. These are the key ingredients for performing a fusion and braiding experiments. The curved chain geometry can also be generalized so that braiding and more complicated manipulations of MZMs can be performed along similar lines as proposed in Ref. 7. However, to preserve the quantum information stored in the MZMs the following conditions have to be satisfied [8]: (i) The time scale of operations $t_0$ has to be much shorter than the tunneling time $t_{\text{tunneling}} \propto e^{L/\zeta_M}$ and thermal excitation time $t_{\text{thermal}} \propto e^{E_{\text{gap}}/k_B T}$, where $L$ is the distance between spatially separated Majoranas (the ones which are not fused intentionally), $\zeta_M$ is the localization length of the Majoranas, $E_{\text{gap}}$ is the minimum excitation energy of the quasiparticles (other than the MZMs) during the braiding cycle, $k_B$ is the Boltzmann constant and $T$ is the temperature. (ii) The time-scale $t_0$ has be much shorter than the quasiparticle poisoning time $t_{\text{poisoning}}$ (which is typically determined by non-equilibrium quasiparticles). (iii) The time-scale $t_0$ should be long compared to $\hbar/E_{\text{gap}}$ to avoid dynamical excitations of the quasiparticles.

The curved chain geometry leads to slowly varying parameters along the chain so that in addition to the MZM there exists low-energy Andreev bound states at the domain wall between nontrivial and trivial regions. This means that $E_{\text{gap}}$ is much smaller than the topological gap $\Delta_{\text{topo}}$ which can be achieved in a linear chain. Thus, also the localization length $\zeta_M$ in the curved chain is much longer than the corresponding length scale $\zeta$ of the linear chain discussed in the main text. Therefore, $t_{\text{tunneling}}$ and $t_{\text{thermal}}$ are much shorter in curved chain geometry than in the linear chain, so that it is challenging to satisfy the requirements $\hbar/E_{\text{gap}} \ll t_0 \ll t_{\text{tunneling}}, t_{\text{thermal}}$. We also point out it is difficult to vary the external magnetic field fast so that it is also difficult to satisfy the requirement $t_0 \ll t_{\text{poisoning}}$.

In the next section we discuss how it is possible to overcome these problems by using small magnets where the magnetization directions are controlled fast using spintronic techniques [9–14].

## Supplementary Note S10.  MAJORANA SPINTRONICS: FUSION AND EXCHANGE OF MAJORANA ZERO MODES

In order to probe the non-Abelian statistics of the MZMs, we propose a tri-junction geometry (see Fig. S18), where Majorana zero modes appear at the interfaces of topologically nontrivial (red) and trivial (black) regimes. The topology of each segment $i = 1, 2, 3$ of the tri-junction (see Fig. S18) is controlled by placing two magnets with magnetizations $M_{iA}$ and $M_{iB}$ on the different sides of the chain. The magnets should be placed within the superconducting coherence length from the impurity sites (in the case of $Sr_2RuO_4$ this is approximately $\sim 70$ nm) and they should be close enough to the surface of the superconductor to cause a magnetic exchange field due to the magnetic proximity effect. If the magnetizations $M_{iA}$ and $M_{iB}$ point in the same direction the segment $i$ realizes approximately an impurity chain in the presence of a homogeneous Zeeman field. On the other hand, if $M_{iA}$ and $M_{iB}$ point in opposite directions, their effect on the impurity bound states cancel each other so that as a good approximation we obtain the Hamiltonian in the absence of Zeeman field. Thus, by designing the magnets so that the magnetizations $M_{iA}$ and $M_{iB}$ have an easy-axis anisotropy along the direction of the segment $i$, the results obtained in the previous sections demonstrate that we can choose the magnitudes of the exchange fields so that the parallel (antiparallel) magnetizations $M_{iA}$ and $M_{iB}$ lead to topologically nontrivial (trivial) phase with large energy gap $E_{\text{gap}}$ and short localization length of MZMs $\zeta_M$. Thus, the MZMs can be robustly manipulated by switching the magnetization directions fast using the spintronic techniques [9–14].

To perform an exchange of MZMs we utilize the anyon teleportation scheme [8, 15, 16], where the exchange of MZMs is obtained via a sequence of projective measurements shown in Fig. S18. According to the universal non-Abelian braiding statistics of the MZMs the exchange of MZMs $\gamma_1$ and $\gamma_2$ is described by applying the unitary operator

$$U = e^{-\frac{\pi}{4}\gamma_1\gamma_2} = \frac{1}{\sqrt{2}}\left(1 + \gamma_2\gamma_1\right) \tag{S47}$$

on the state of the system [8]. The teleportation scheme is based on the decomposition[8, 15, 16]

$$\Pi_{03}\Pi_{02}\Pi_{01}\Pi_{03} = \sqrt{\frac{1}{8}}\Pi_{03} \otimes \frac{1}{\sqrt{2}}(1 + \gamma_2\gamma_1), \tag{S48}$$

where each operator

$$\Pi_{kl} = \frac{1}{2}(1 + i\gamma_k\gamma_l) \tag{S49}$$

describes a projective measurement of the parity $P_{kl} = i\gamma_k\gamma_l$ of MZMs $\gamma_k$ and $\gamma_l$ with the outcome of the measurement being +1. Therefore, based on Eqs. (S47) and (S48), it is clear that the exchange of MZMs $\gamma_1$ and $\gamma_2$ can be performed as follows: (i) Initialize the parity $P_{03} = 1$ by fusing MZMs $\gamma_0$ and $\gamma_3$, performing a projective measurement of $i\gamma_0\gamma_3$ and continuing only if the outcome of the measurement is +1. (ii) Perform similarly a measurement of the parity

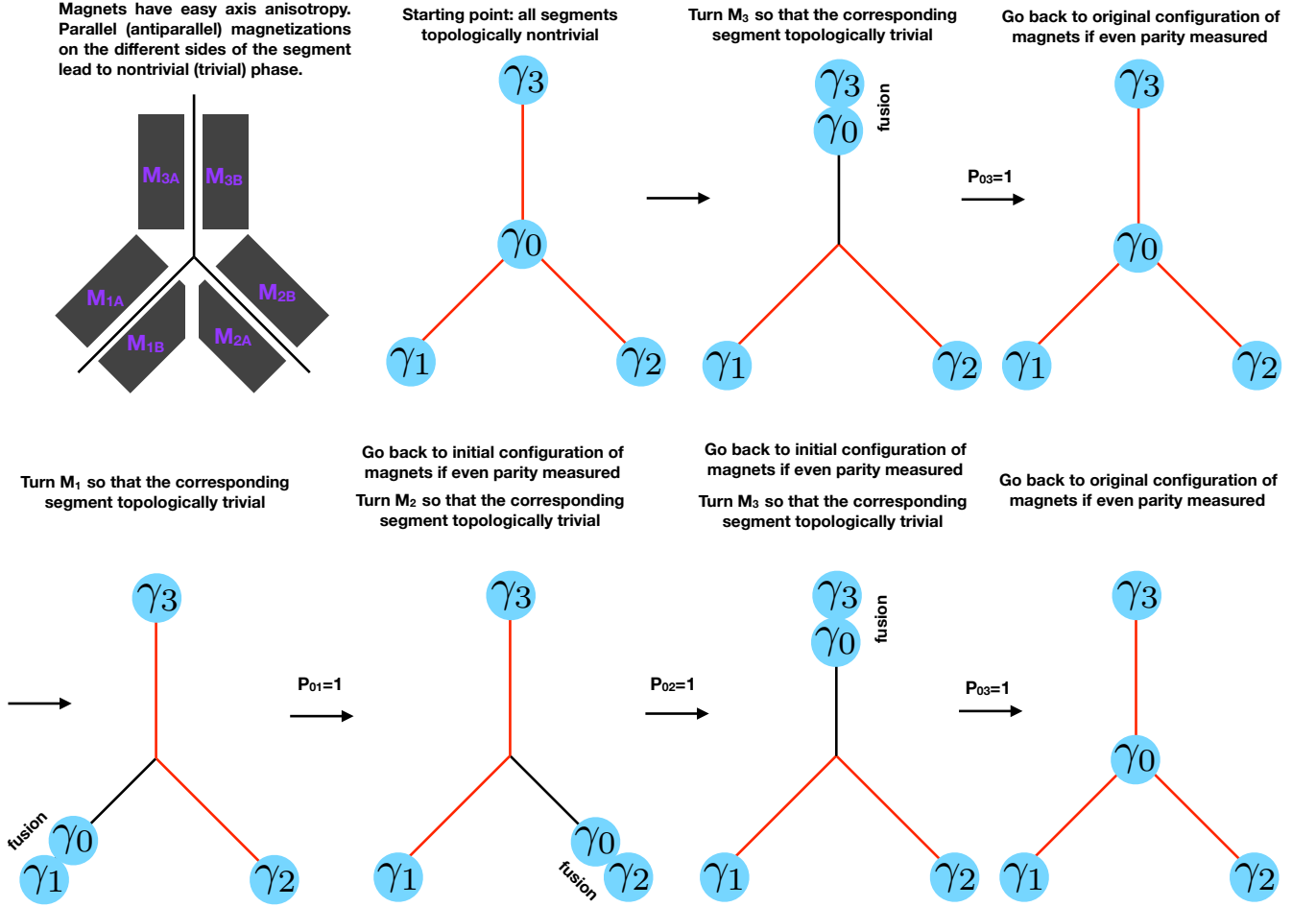

FIG. S18. Exchange of MZMs by utilizing the anyon teleportation scheme. The unitary operation describing the non-Abelian Majorana braiding statistics [Eq. (S47)] can be decomposed into four successive projective measurements of parity operators [Eq. (S48)]. Each of these parity operators $i\gamma_0\gamma_i$ can be measured by rotating the magnetizations $M_{iA}$ and $M_{iB}$ from parallel to antiparallel configuration resulting in hybridization of $\gamma_0$ and $\gamma_i$ so that their parity can be measured using a charge sensor.

$i\gamma_0\gamma_1$ and continue if the outcome of the measurement is $+1$. (iii) Perform a measurement of the parity $i\gamma_0\gamma_2$ and continue if the outcome of the measurement is $+1$. (iv) Perform a measurement of the parity $i\gamma_0\gamma_3$. The exchange of Majoranas $\gamma_1$ and $\gamma_2$ is successfully executed if the outcome of the measurement is $+1$. In Fig. S18 we illustrate how all the steps of this process can be realized with the help of the small magnets. This exchange operation has a probabilistic element because in each projective measurement the outcome of the measurement can also be $-1$. Therefore, one typically needs to repeat the process many times to successfully execute the exchange operation.

Finally, we point out that the circuit shown in Fig. S18 can also be utilized to perform the fusion experiment discussed in the main text without the obstacles mentioned in the previous section. Moreover, the ideas presented in this section can be generalized for a construction of a fully scalable network of impurity chains and small magnets, where the MZMs are manipulated by spintronic means.


\fi

\end{document}